\documentclass[smallcondensed]{article}
\def\fullversion{}
\pdfoutput=1
\usepackage[utf8]{inputenc}
\usepackage{enumerate,ifthen,xspace}
\usepackage{substr}
\usepackage{graphicx,color}
\usepackage{amsmath,amsfonts,amstext,amssymb,mathtools}
\usepackage{amsbsy}
\makeatletter\@ifclassloaded{svjour3}{}{\usepackage{amsthm}}\makeatother
\usepackage{stmaryrd}
\usepackage{caption}
\usepackage{subfig}
\usepackage{sgamevar}
\usepackage{url}
\usepackage[pdfusetitle,bookmarks=false]{hyperref}
\usepackage[round]{natbib}
\let\cite\citep
\usepackage[ruled]{algorithm2e}
\usepackage{aliascnt}
\usepackage{cleveref}
\usepackage{paralist}
\usepackage{times}

\ifx\crefformat\undefined
\newcommand{\crefformats}[7]{}
\else
\newcommand{\crefformats}[7]{%
  \crefformat{#3}{##2\ifthenelse{\equal{#4}{}}{}{#4~}#1##1#2##3}
  \Crefformat{#3}{##2\ifthenelse{\equal{#5}{}}{}{#5~}#1##1#2##3}
  \crefrangeformat{#3}{\ifthenelse{\equal{#6}{}}{}{#6~}##3#1##1#2##4--##5#1##2#2##6}
  \Crefrangeformat{#3}{\ifthenelse{\equal{#7}{}}{}{#7~}##3#1##1#2##4--##5#1##2#2##6}
  \crefmultiformat{#3}{\ifthenelse{\equal{#6}{}}{}{#6~}##2#1##1#2##3}{ and~##2#1##1#2##3}{, ##2#1##1#2##3}{ and~##2#1##1#2##3}
  \Crefmultiformat{#3}{\ifthenelse{\equal{#7}{}}{}{#7~}##2#1##1#2##3}{ and~##2#1##1#2##3}{, ##2#1##1#2##3}{ and~##2#1##1#2##3}
  \crefrangemultiformat{#3}{\ifthenelse{\equal{#6}{}}{}{#6~}##2#1##1#2##3}{ and~##2#1##1#2##3}{, ##2#1##1#2##3}{ and~##2#1##1#2##3}
  \Crefrangemultiformat{#3}{\ifthenelse{\equal{#7}{}}{}{#7~}##2#1##1#2##3}{ and~##2#1##1#2##3}{, ##2#1##1#2##3}{ and~##2#1##1#2##3}
}

\makeatletter
\@ifclassloaded{llncs}{
  \crefformats{}{}{section}{Sect.}{Section}{Sects.}{Sections}
  \crefformats{}{}{appendix}{App.}{Appendix}{App.}{Appendices}
  \crefformats{}{}{subsection}{Sect.}{Section}{Sects.}{Sections}
  \crefformats{}{}{subsubsection}{Sect.}{Section}{Sects.}{Sections}
  \crefformats{}{}{chapter}{Chap.}{Chapter}{Chaps.}{Chapters}
  \crefformats{}{}{part}{Part}{Part}{Parts}{Parts}
  \crefformats{}{}{figure}{Fig.}{Figure}{Figs.}{Figures}
  \crefformats{}{}{table}{Table}{Table}{Tables}{Tables}
}{
   \crefformats{}{}{section}{Section}{Section}{Sections}{Sections}
   \crefformats{}{}{appendix}{Appendix}{Appendix}{Appendices}{Appendices}
   \crefformats{}{}{subsection}{Section}{Section}{Sections}{Sections}
   \crefformats{}{}{subsubsection}{Section}{Section}{Sections}{Sections}
   \crefformats{}{}{chapter}{Chapter}{Chapter}{Chapters}{Chapters}
   \crefformats{}{}{part}{Part}{Part}{Parts}{Parts}
   \crefformats{}{}{figure}{Figure}{Figure}{Figures}{Figures}
   \crefformats{}{}{table}{Table}{Table}{Tables}{Tables}
}
\makeatother
\crefformats{}{}{footnote}{footnote}{footnote}{footnotes}{footnotes}
\fi

\ifx\finalversion\undefined
\newcommand{\marginlabel}[2]{%
  \mbox{}%
  \marginpar[\raggedleft\hspace{0pt}#1]{\raggedright\hspace{0pt}#2}%
}
\ifx\tikzstyle\undefined
\newcommand{\todoar}[2][]{\todo[#1]{#2}}
\else\tikzstyle{todoarrow}=[opacity=0.4,gray,-stealth]
\newcommand{\todoar}[2][]{%
  \marginlabel{\small #2}
        {\tikz[remember picture,overlay,baseline=(todoarrowstart.220)]\node(todoarrowstart){};$\lhd$ \small #2}%
  \ifthenelse{\equal{#1}{}}{}{{\color{red}[}#1{\color{red}]}}%
  \tikz[remember picture,overlay]\node[inner sep=2pt](todoarrowend){};%
  \tikz[remember picture,overlay]\path(todoarrowstart)edge[todoarrow,out=190,in=-45](todoarrowend);%
}
\fi
\newcommand{\todo}[2][]{%
  \marginlabel{$\rhd$ {\small #2}}{$\lhd$ \small #2}%
  \ifx\color\undefined%
  \ifthenelse{\equal{#1}{}}{}{{[}#1{]}}%
  \else%
  \ifthenelse{\equal{#1}{}}{}{{\color{red}[}#1{\color{red}]}}%
  \fi%
}
\fi

\newcounter{autoexternalpgf}
\newcommand{\abs}[1]{\lvert#1\rvert}
\newcommand{\suchthat}{\,|\,}
\newenvironment{mainclaim}{\begin{center}}{\end{center}}

\makeatletter
\@ifclassloaded{beamer}{}{%
  \newcommand{\newtheoremwithalias}[3]{%
    \ifx\newaliascnt\undefined
    \newcounter{#1}
    \else
    \newaliascnt{#1}{#2}
    \fi
    \newtheorem{#1}[#1]{#3}
    \ifx\aliascntresetthe\undefined\else
    \aliascntresetthe{#1}
    \fi
  }
  \ifx\creflastconjunction\undefined\else
  \renewcommand{\creflastconjunction}{ and }
  \fi

  \@ifclassloaded{llncs}{
    \if@envcntsame\errmessage{cleveref naming doesn't work because no aliascntrs used in llncs.cls}
    \else\if@envcntsect
    \newtheorem{observation}{Observation}[section]
    \newtheorem{fact}{Fact}[section]
    \else

    \fi\fi
    \newcommand{\qedhere}{\qed}
    \crefformats{(}{)}{equation}{}{Equation}{}{Equations}
  }{
    \newtheorem{theorem}{Theorem}[section]
    \newtheoremwithalias{lemma}{theorem}{Lemma}
    \newtheoremwithalias{proposition}{theorem}{Proposition}
    \newtheoremwithalias{corollary}{theorem}{Corollary}
    \newtheoremwithalias{observation}{theorem}{Observation}
    \newtheoremwithalias{fact}{theorem}{Fact}
    \newtheoremwithalias{conjecture}{theorem}{Conjecture}

    \ifx\theoremstyle\undefined\else\theoremstyle{remark}\fi
    \newtheoremwithalias{numberednote}{theorem}{Note}
    \newtheoremwithalias{numberedremark}{theorem}{Remark}

    \ifx\theoremstyle\undefined\else\theoremstyle{definition}\fi
    \newtheoremwithalias{definition}{theorem}{Definition}
    \newtheoremwithalias{example}{theorem}{Example}

    \crefformats{(}{)}{equation}{Equation}{Equation}{Equations}{Equations}

  }

  \crefformats{}{}{theorem}{Theorem}{Theorem}{Theorems}{Theorems}
  \crefformats{}{}{lemma}{Lemma}{Lemma}{Lemmas}{Lemmas}
  \crefformats{}{}{proposition}{Proposition}{Proposition}{Propositions}{Propositions}
  \crefformats{}{}{corollary}{Corollary}{Corollary}{Corollaries}{Corollaries}
  \crefformats{}{}{observation}{Observation}{Observation}{Observations}{Observations}
  \crefformats{}{}{fact}{Fact}{Fact}{Facts}{Facts}
  \crefformats{}{}{conjecture}{Conjecture}{Conjecture}{Conjectures}{Conjectures}
  \crefformats{}{}{numberednote}{Note}{Note}{Notes}{Notes}
  \crefformats{}{}{numberedremark}{Remark}{Remark}{Remarks}{Remarks}
  \crefformats{}{}{definition}{Definition}{Definition}{Definitions}{Definitions}
  \crefformats{}{}{example}{Example}{Example}{Examples}{Examples}
}
\makeatother

\makeatletter\@ifclassloaded{svjour3}{
  \newcommand{\proofof}{}
  \renewcommand{\qedhere}{} 
  \smartqed
  \newcommand{\qedhack}{\vskip-4mm~}
  \newcommand{\oldendproof}{}
  \let\oldendproof\endproof
  \renewcommand{\endproof}{\qed\oldendproof}
}{
  \newcommand{\qedhack}{}
  \newcommand{\proofof}{Proof of }
}\makeatother

\newcommand{\oldendexample}{}
\let\oldendexample\endexample
\renewcommand{\endexample}{\qed\oldendexample}

\allowdisplaybreaks[1]

\crefformats{}{}{algocf}{Algorithm}{Algorithm}{Algorithms}{Algorithms}

\ifx\hypersetup\undefined\else\hypersetup{pdftitle={Distributed iterated elimination of strictly dominated strategies}}\fi
\title{Distributed iterated elimination of strictly dominated strategies%
\thanks{The first and third authors were supported by a GLoRiClass fellowship
funded by the European Commission (Early Stage Research Training
Mono-Host Fellowship MEST-CT-2005-020841).
Most of this work was done at the Institute for Logic, Language and Computation, University~of~Amsterdam, Netherlands.}
\ifx\fullversion\undefined%
\protect\footnote{Proofs are omitted for space reasons. The full version is available at
  \url{http://arxiv.org/abs/0908.2399v1}.}%
\fi}
\ifx\hypersetup\undefined\else\hypersetup{pdfauthor={Andreas Witzel and Krzysztof R. Apt and Jonathan A. Zvesper}}\fi

\makeatletter\@ifclassloaded{svjour3}{

\author{Andreas Witzel \and Krzysztof R. Apt \and Jonathan A. Zvesper}

\institute{University~of~Amsterdam, Science Park~904, 1098XH~Amsterdam
    \and
    CWI, Kruislaan~413, 1098SJ~Amsterdam, The~Netherlands}

\institute{
  A. Witzel \at
  CIMS, New York University,
  715 Broadway, Room 1010,
  New York, NY 10003, USA\\
  \email{awitzel@nyu.edu}
  \and
  K.R. Apt \at
  CWI, Science Park 123,
  1098 XG Amsterdam,
  The Netherlands\\
  \email{apt@cwi.nl}
  \and
  J.A. Zvesper \at
  Oxford University Computing Laboratory,
  Room 213, Wolfson Building, Parks Road,
  Oxford OX1 3QD, United Kingdom
}

}{
\author{
Andreas Witzel
\and
Krzysztof R. Apt
\and
Jonathan A. Zvesper
\\[2ex]
ILLC, University of Amsterdam, the Netherlands \\
and CWI, Amsterdam}
}\makeatother

\ifx\Newassociation\undefined

\newenvironment{MovedProof}[1]{}{}

\newenvironment{movedProof}{%
  \catcode`\{=12
  \catcode`\}=12
  \DoNull
}{%

  \ignorespacesafterend

}

\begingroup
  \catcode`\(=1
  \catcode`\)=2
  \catcode`\{=12
  \catcode`\}=12
  \def\x(%
    \endgroup
    \newcommand\DoNull()
    \long\def\DoNull##1\end{movedProof}(\end(movedProof))
  )
\x
\else
\Newassociation{movedProof}{MovedProof}{movedProofs}
\renewenvironment{MovedProof}[1]{}{}
\Opensolutionfile{movedProofs}
\fi

\newcommand{\dfn}[2][***gurgelblubbla]{%
\ifthenelse{\equal{#1}{***gurgelblubbla}}{\index{#2}}%
{\ifthenelse{\equal{#1}{}}{}{\index{#1}}}%
{\bfseries #2}}
\newcommand{\dfnless}[2][***gurgelblubbla]{%
\ifthenelse{\equal{#1}{***gurgelblubbla}}{\index{#2}}%
{\ifthenelse{\equal{#1}{}}{}{\index{#1}}}%
\emph{#2}}
\newcommand{\mpunct}{}
\newcommand{\qinf}[1]{``#1''\xspace}

\newcommand{\G}{\mathcal{G}}
\newcommand{\group}{A}
\newcommand{\hyarc}{A}

\newcommand{\agents}{N}

\newcommand{\sdglobal}{\ensuremath{sd^g}\xspace}
\newcommand{\sdlocal}{\ensuremath{sd}\xspace}

\newcommand{\mvee}{\text{ or }}
\newcommand{\mexists}{\exists}
\newcommand{\mforall}{\forall}
\renewcommand{\iff}[1][***gurgelblubbla]{\ifthenelse{\equal{#1}{***gurgelblubbla}}{\text{ iff }}{%
\text{ iff }&\ifthenelse{\equal{#1}{}}{}{\text{\it(#1)}}\\
}}
\newcommand{\ifff}[1]{%
\text{ iff }&\text{\it(#1)}\\
}

\newcommand{\domin}{\mathrm{domin}}
\newcommand{\dominated}{\mathrm{dom}}

\newcommand{\langposk}{\mathcal{L}^+}

\newcommand{\init}{V} 
\newcommand{\msgs}{\ensuremath{M}\xspace}

\newcommand{\allmsgs}[1]{\ensuremath{\msgs^{\mathrm{all}}_{#1}}\xspace}
\newcommand{\prf}[3]{#2_{#1}#3\succ_{\kern-.5pt#2_{-#1}}#2_{#1}}
\newcommand{\pref}[4]{#3_{#1}\succ_{\kern-.5pt#2_{-#1}}#4_{#1}}
\newcommand{\preferred}[3]{#2\succ_{#1}#3}

\newcommand{\prefeq}[4]{#3_{#1}\succeq_{\kern-.5pt#2_{-#1}}#4_{#1}}
\newcommand{\preferredeq}[3]{#2\succeq_{#1}#3}

\renewcommand{\vee}{\lor}

\newcommand{\state}[1][]{\ensuremath{(V#1,M#1)}\xspace}

\newcommand{\msg}[3]{\ensuremath{(#1,#2,#3)}\xspace}
\newcommand{\bits}{\ensuremath{\mathrm{At}}\xspace}
\newcommand{\atoms}{\bits}
\newcommand{\knows}[1]{\ensuremath{K_{#1}}\xspace}
\newcommand{\ck}[1]{\ensuremath{C_{#1}}\xspace}

\newcommand{\restr}{\kern-3pt\restriction}

\ifx\pgfversion\undefined
  \input{pgfexternal.tex}
\else

  \usetikzlibrary{positioning}
  \usetikzlibrary{automata}
  \usetikzlibrary{shapes}
  \usetikzlibrary{shapes.callouts}

  \pgfdeclarelayer{background}
  \pgfsetlayers{background,main}

  \tikzstyle{privateknowledge}=[draw,font=\scriptsize,inner sep=1.5pt,fill=white,node distance=4mm,cloud callout,cloud ignores aspect,cloud puffs=14,cloud puff arc=100]
  \tikzstyle{commonknowledge}=[draw,font=\scriptsize,inner sep=1pt,fill=white,node distance=1mm,ellipse callout]
  \tikzstyle{nextpic}=[bend right,ultra thick,opacity=.6,->,dashed,shorten >=5mm,shorten <=5mm]

  \tikzstyle{vx}=[circle, draw, inner sep=1pt, minimum width=10pt]
  \tikzstyle{ed}=[->,>=stealth]
  \tikzstyle{ed2}=[ed,<->]
  \tikzstyle{mapping}=[gray,opacity=0.8,semithick] 
  \tikzstyle{bgcircle}=[circle,fill=gray,opacity=.3,outer sep=6pt]
  \tikzstyle{scaled}=[] 

  \usetikzlibrary{calc,fit}
  \tikzstyle{commgraphnode}=[circle]
  \tikzstyle{commgraphedge}=[]
  \tikzstyle{commgraphhyperarc}=[ellipse,fill=gray,opacity=.3,inner sep=0pt]

  \ifthenelse{\lengthtest{\pgfversion cm<1.18cm}}{
  }{
    \ifx\regeneratepgf\undefined\pgfrealjobname{\jobname}
    \else\pgfrealjobname{\regeneratepgf}
    \fi
  }

\fi

\begin{document}

\maketitle

\begin{abstract}
  We characterize epistemic consequences of truthful communication among rational agents in
  a game-theoretic setting.
  To this end we introduce normal-form games equipped with an
  interaction structure, which specifies which groups of players can
  communicate their preferences with each other.  We then focus on a
  specific form of interaction, namely a distributed form of iterated
  elimination of strictly dominated strategies (IESDS), driven by
  communication among the agents.  
We study the outcome of IESDS after some (possibly all) messages about
players' preferences have been sent.  The main result of the paper,
\cref{result:alg-equiv-formula}, provides an epistemic justification
of this form of IESDS.
\ifx\keywords\undefined\else\fi
\end{abstract}

\section{Introduction}
\label{sec:intro-stratelim}

\subsection{Motivation and framework}
\label{sec:motivation}

One of the main topics in the area of multiagent reasoning is study of
epistemic reasoning of \emph{communicating} agents in a distributed
setting, see, e.g., \cite{fagin_knowledge-based_1997,chandy_processes_1986}.
Game theory focuses on interaction between different types of agents,
namely \emph{rational} agents, whose objective is to maximize their
utility.

Our interest is to analyze epistemic reasoning of agents who exhibit
both characteristics, i.e. who communicate and are rational.  We
assume that such agents are involved in \emph{distributed decision
making}, by which we mean an interactive process during which agents
repeatedly combine their local information with new information
obtained through communication with other agents in order to arrive at
a (possibly common) conclusion by means of a deductive process.

To make such a general study meaningful we focus on a specific
instance of distributed decision making, namely iterated elimination
of strictly dominated strategies (IESDS, \cite{fudenberg_game_1991}) driven by the
acquisition of new information through communication.  In our setup
each agent repeatedly combines local information about his
preference among his strategies with new information acquired through
interaction with other agents. This allows each agent to increasingly
eliminate more strategies.  We are interested in characterizing
knowledge of each agent at the end of this decision making process.


To formalize this setting we introduce a game-theoretic framework which combines
\emph{locality} and \emph{interaction}.  We assume a setting of imperfect information 
in which the players' preferences are \emph{not} commonly known.  Instead, the initial
information of each player only covers \emph{his own} preferences, and
the players can truthfully \emph{communicate} this information in the
fixed groups to which they belong.  So locality refers to
the \emph{information} about preferences and interaction refers to
\emph{communication} within (possibly overlapping) groups of players.

To realize this framework we augment a normal-form game with an
\emph{interaction structure} of \cite{apt_knowledge_2009} that consists
of (possibly overlapping) groups of players within which synchronous
communication is possible.

More precisely, we make the following assumptions:
\begin{itemize}
\item the players initially know 
the interaction structure and their own preferences,
\item they are rational, in the sense that they would not play a strictly dominated strategy,
\item they   can communicate atomic information about their preferences
within any group they belong to,
\item once a message is communicated, it is commonly known within the group,
\item communication is truthful and synchronous,
\item the players have no knowledge other than what follows from these assumptions,
\item the above assumptions are common knowledge.
\end{itemize}

In our communication setting, intuitively, information can be either unknown
to some player or commonly known in a group. Acting on partial
information thus necessitates to figure out precisely which
information is commonly known in a group.
Note that in this context strategic (i.e., possibly untruthful)
communication conveys no conclusive information. Consequently, study of
epistemic consequences of such communication requires
additional assumptions, like probability distributions over statements and meanings,
and can be carried out only after the setting with truthful communication
has been investigated.
We therefore focus on truthful communication and return briefly to this matter in \cref{sec:conclusions-stratelim}.

The following example illustrates the general procedure that we examine.

\begin{example}
  \label{ex:trace-stratelim}
  Assume three players $1,2,3$ (row, column, and matrix player)
  who can only communicate pairwise
  and who play the following game,
  where ordinal preferences are represented by numerical payoffs:

  \hfill
  \begin{game}{2}{2}[$A$]
    \> $L$   \> $R$   \\
    $U$ \> $0,1,0$ \> $0,0,1$ \\
    $D$ \> $1,0,0$ \> $1,1,1$~
  \end{game}
  \hfill
  \begin{game}{2}{2}[$B$]
    \> $L$ \> $R$ \\
    $U$ \> $0,1,1$ \> $0,0,0$ \\
    $D$ \> $1,0,1$ \> $1,1,0$~
  \end{game}
  \hfill~

  Note that the payoffs of players 1 and 2 do not depend on the strategy 
  on player~3 and the payoff of player~3 depends only on the strategy of player~2.
  This game can easily be solved by iterated elimination of strictly dominated strategies,
  yielding the profile $(D,R,A)$.
  However, assuming that each player is ignorant about the other players' preferences,
  the initial situation looks as follows
  from the perspective of players 1, 2, and 3, respectively:

  \hfill
  \begin{game}{2}{2}[$A$]
    \> $L$   \> $R$   \\
    $U$ \> $0,.,.$ \> $0,.,.$ \\
    $D$ \> $1,.,.$ \> $1,.,.$~
  \end{game}
  \hfill
  \begin{game}{2}{2}[$B$]
    \> $L$ \> $R$ \\
    $U$ \> $0,.,.$ \> $0,.,.$ \\
    $D$ \> $1,.,.$ \> $1,.,.$~
  \end{game}
  \hfill~

  \hfill
  \begin{game}{2}{2}[$A$]
    \> $L$   \> $R$   \\
    $U$ \> $.,1,.$ \> $.,0,.$ \\
    $D$ \> $.,0,.$ \> $.,1,.$~
  \end{game}
  \hfill
  \begin{game}{2}{2}[$B$]
    \> $L$ \> $R$ \\
    $U$ \> $.,1,.$ \> $.,0,.$ \\
    $D$ \> $.,0,.$ \> $.,1,.$~
  \end{game}
  \hfill~

  \hfill
  \begin{game}{2}{2}[$A$]
    \> $L$   \> $R$   \\
    $U$ \> $.,.,0$ \> $.,.,1$ \\
    $D$ \> $.,.,0$ \> $.,.,1$~
  \end{game}
  \hfill
  \begin{game}{2}{2}[$B$]
    \> $L$ \> $R$ \\
    $U$ \> $.,.,1$ \> $.,.,0$ \\
    $D$ \> $.,.,1$ \> $.,.,0$~
  \end{game}
  \hfill~

  Given player~1's rationality,
  he will not play~$U$, as it is strictly dominated by $D$.
  However, in the current situation,
  neither of the other two players can exclude player~1's playing $U$,
  given the information at their disposal.

  Assume now that player~1 communicates his preferences to player~2,
  thus creating common knowledge about them (among himself and player~2).
  Player~2's picture then looks as follows:

  \hfill
  \begin{game}{2}{2}[$A$]
    \> $L$   \> $R$   \\
    $U$ \> $0,1,.$ \> $0,0,.$ \\
    $D$ \> $1,0,.$ \> $1,1,.$~
  \end{game}
  \hfill
  \begin{game}{2}{2}[$B$]
    \> $L$ \> $R$ \\
    $U$ \> $0,1,.$ \> $0,0,.$ \\
    $D$ \> $1,0,.$ \> $1,1,.$~
  \end{game}
  \hfill~

  Player~2, knowing that player~1 is rational, can conclude that player~1 will not play~$U$.
  Consequently player~2 can remove his own strategy $L$ from consideration,
  since with player~1 playing $D$, $L$ is dominated by $R$.
  If player~2 now communicates his preferences both to players 1 and 3,
  then player~1 will be able to conclude that player~2 will not play~$L$;
  however, player~3 will not be able to conclude this,
  since he lacks the information that player~2 knows that player~1 will not play~$U$.
  Given that communication is only about players' own preferences,
  and given that in this example we only allowed pairwise communication,
  player~3 will not be able to deduce that player~2 will play~$R$,
  and consequently will not be able to eliminate his strategy $B$.
\end{example}
In this paper, we look at settings of restricted communications like in this example,
and characterize exactly what strategies can be safely eliminated at different stages of communication.


\subsection{Results}
\label{sec:results}

We study the outcome of iterated elimination of strictly dominated
strategies (IESDS) in the above setting in any state of partial
communication (i.e., where some messages about players' preferences
have been sent), and in particular in the state of full communication,
where all communication permitted by the interaction structure has
taken place.  In terms of distributed decision making, local
information consists of players' preferences and new information
consists of information about other players' preferences obtained
through the received messages.  In turn, deduction consists of the
elimination of strategies, and the conclusion is the outcome of IESDS.

We use the results from our previous work~\cite{apt_knowledge_2009} to
prove that this outcome realizes an epistemic formula that describes
what the players know in the considered partial communication state.

\subsection{Background and related work}
\label{sec:background}

Our epistemic analysis belongs to a large body of research within
  game theory concerned with the study of players' \emph{knowledge}
  and \emph{beliefs}, see, e.g.~\cite{battigalli_recent_1999}.  In
  particular, \cite{BD87} and \cite{tan_bayesian_1988} have shown that
  if the payoff functions are commonly known and the players are
  \dfnless{rational} and have common knowledge of each other's
  rationality, they will only play strategies that survive IESDS,
  where one uses strict dominance by a mixed strategy.  
  
  When one restricts one's attention to pure strategies, then, as
  shown in \cite{AZ10}, the above implication holds for arbitary
  games.  It is also clarified there that the corresponding
  implication does not hold for weak dominance, even for finite
  two-player games.
  
  The mathematical reason is that the \emph{global} version of strict
  dominance, $sd^{g}$, that we introduce in \cref{sec:epist-form-strict} is
  monotonic, while the global version of weak dominance is \emph{not}.
  Now, our approach crucially depends on the monotonicity of $sd^{g}$,
  which explains why the results of our paper do not carry over to
  weak dominance.
  
  The concept of IESDS is closely related to that of
  \emph{rationalizability} of \cite{Ber84,Pea84} that focuses on the
  strategies that survive iterated elimination of never best
  responses.  In fact, (see, e.g.~\cite[Proposition 61.2]{OR94}) the
  concepts of a strict dominance by a \emph{mixed} strategy and of a
  never best response to a \emph{correlated} strategy coincide.
  However, we use here only pure strategies, and consequently in our
  setup these concepts differ.
  
  Further, as shown in \cite{vB07}, for finite games the outcome of
  IESDS can be characterized using the concept of a \emph{public
    announcement} due to \cite{Pla89}.  In \cite{AZ10a} we generalized
  this characterization to infinite games and clarified that it holds
  for monotonic dominance notions, including the global version of
  strict dominance, but excluding the customary strict dominance.
  
  Our framework stresses the locality of information about preferences
  in combination with communication and consequently leads to a
  different epistemic analysis.  In particular, in our setting the
  analysis of players' knowledge requires taking into account players'
  reasoning about group communication.


It is useful to clarify the difference between our framework and
\dfnless[game!graphical]{graphical games} of~\cite{kearns_graphical_2001}.
In these games 
a locality assumption is formalized by assuming a graph structure
over the set of players and using payoff functions which depend only
on the strategies of players' neighbors.
%
The absence of communication precludes a distributed view of IESDS.

\subsection{Plan of the paper}

In \cref{sec:preliminaries}, we review the basic notions concerning
normal-form games, strict dominance and operators on the restrictions
of games.  Next, in \cref{sec:iter-strat-elim}, we study the outcome
of IESDS in the presence of an interaction structure.  We first look
at the outcome resulting after all communication permitted in the
given interaction structure has taken place, and then consider the
outcome obtained in an arbitrary state of partial communication.

The connection with knowledge is made in \cref{sec:epist-found}, where
we provide in this setting the epistemic characterization of 
IESDS.
Finally, in \cref{sec:conclusions-stratelim}, we suggest some future research directions.

In the Appendix we provide the omitted proofs.
An initial, short version of this paper appeared as \cite{WAZ09}.
\section{Preliminaries}
\label{sec:preliminaries}

Following \citet{OR_GT}, a \dfn{normal-form game}
(in short, \dfn{game}) for players $\agents = \{1, \dots, n\}$ (with $n > 1$) is a tuple
$ 
(S_1, \dots, S_n, \succsim_1, \dots, \succsim_n),
$ 
where for each $i \in\agents$,
\begin{itemize}
\item $S_i$ is the non-empty, finite set of \dfn[strategy]{strategies} 
available to player~$i$.
We write $S$ to abbreviate the set of \dfn[strategy!profile]{strategy profiles}:
$S=S_1\times\dots\times S_n$.
\item $\succsim_i$ is the
\dfn[]{preference relation} (i.e., a complete transitive reflexive binary relation) for player~$i$, 
so $\succsim_i \subseteq S \times  S$.
\end{itemize}

Since we are considering strict dominance, we will use from now on
only the strict counterparts of the preference relations, denoted by
$\succ_i$.  In this qualitative approach one cannot introduce mixed
strategies. However, they are not used in our approach.  The extension
of our results to the iterated elimination of strict dominance by a
mixed strategy is not obvious. The reason is that in the proofs we
crucially rely on the fact that one can form a finite disjunction over
the set of strictly dominating strategies.  In specific examples we
shall use the customary payoff functions.

As usual we denote player $i$'s strategy in a strategy profile $s\in S$ by $s_i$,
and the tuple consisting of all other strategies by $s_{-i}$,
i.e.,
$ 
s_{-i}=(s_1, \dots, s_{i-1}, s_{i+1}, \dots, s_n)\mpunct.
$ 
Similarly, we use $S_{-i}$ to denote $S_1\times\dots \times S_{i-1}\times S_{i+1}\times\dots\times S_n$,
and for $s_i'\in S_i$ and $s_{-i}\in S_{-i}$ we write $(s'_i, s_{-i})$
to denote $(s_1, \dots, s_{i-1}, s'_i, s_{i+1}, \dots, s_n)$.
Finally, we use $\prf is'$ as a shorthand for $(s_i',s_{-i})\succ_i(s_i,s_{-i})$.

In the subsequent considerations each considered game is identified
with the set of statements of the form $\prf is'$, where $i \in N$,
$s_i, s'_i \in S_i$, and $s_{-i} \in S_{-i}$ and both strict dominance notions are formulated in terms
of such statements. Therefore the
results of this paper apply equally well to the \dfn{strategic games
  with parametrized preferences} introduced in
\cite{ARV08}. In these games instead of the
preferences relations $\succ_1, \dots, \succ_n$, for each joint
strategy $s_{-i}$ of the opponents of player $i$ a strict preference
relation $\succ_{s_{-i}}$ over the strategies of player $i$ is given.
So in strategic games with parametrized preferences players cannot
compare two arbitrary strategy profiles.




Fix now an \emph{initial}  game
$\G := (S_1, \dots, S_n, \succ_1, \dots, \succ_n)$.
We say that the tuple $(S'_1, \dots, S'_n)$ is a
\dfn[restriction!of a game]{restriction} of $\G$ if
each $S'_i$ is a subset of $S_i$.
We identify the restriction $(S_1, \dots, S_n)$ with $\G$.

The restrictions of $\G$ are ordered by the component-wise set inclusion:
\[
\mbox{$(G_1, \dots, G_n) \subseteq (G'_1, \dots, G'_n)$ iff $G_i \subseteq G'_i$ for all $i \in \{1, \dots, n\}$.}
\]

To analyze iterated elimination of strategies from the initial game
$\G$, we view such procedures as operators on the finite lattice formed
by the set of restrictions of $\G$ ordered by the above inclusion
relation.


For any restriction $\G' := (S'_1, \dots, S'_n)$ of $\G$ and strategies $s_i,s_i' \in S_i$,
we say that $s_i$ is \dfn[strategy!dominated]{strictly dominated by $s_i'$ on $S'_{-i}$},
and write $\preferred{S_{-i}'}{s_i'}{s_i}$,
if $\preferred{s_{-i}'}{s_i'}{s_i}$ for all $s_{-i}'\in S_{-i}'$.
We use
\[
\sdlocal(s_i, \G')\text{ to abbreviate }\neg \mexists s'_i
  \in S'_i \: \mforall s_{-i}'\in S_{-i}' \: s'_i \succ_{s_{-i}'} s_i.
\]
  That is, $\sdlocal(s_i, \G')$ holds iff strategy $s_i$ of
  player $i$ survives elimination of strictly dominated strategies with respect to $S'_i$, i.e.,
  iff it is not strictly dominated on $S'_{-i}$ by any strategy from $S'_i$.


%
%
%

Given an operator $T$ on a finite lattice $(D, \subseteq)$ with the
largest element $\top$, $X \in D$, and $k \geq0$, we denote by
$T^{k}(X)$ the $k$-fold iteration of $T$ starting at $X$, so with $T^0(X) = X$,
and put $T^{\infty}(X) := \bigcap_{k \geq 0} T^k(X)$, where $\bigcap$ is the meet operation
entailed by $\subseteq$.  We abbreviate
$T^{\alpha}(\top)$ to $T^{\alpha}$.

We call $T$ \dfn[monotonic!operator]{monotonic} if for all $X, Y \in D$,
we have that
$ 
X \subseteq Y\text{ implies }T(X) \subseteq T(Y),
$ 
and \dfn[contracting!operator]{contracting} if for all $X \in D$
we have that $T(X) \subseteq X$.





Finally, as in \cite{apt_knowledge_2009}, an \dfn{interaction structure} $H$
is a \dfnless{hypergraph} on $N$, i.e.,
a set of non-empty subsets of $\agents$, called \dfnless[]{hyperarcs}.

\section{Iterated strategy elimination}
\label{sec:iter-strat-elim}
In this section we define procedures for iterated elimination of strictly dominated strategies.
Let us fix a  game $\G=(S_1, \dots, S_n, \succ_1, \dots, \succ_n)$
for players $\agents$ and an interaction structure $H\subseteq 2^{\agents}\setminus\{\emptyset\}$.
In \cref{sec:static-setting}, we look at the outcome reached
after all communication permitted by~$H$ has taken place,
that is, when within each hyperarc of~$H$ all of its members' preferences have been communicated.
In \cref{sec:intermediate-states}, we then look
at the outcomes obtained in any particular state of partial communication.

The formulations we give here make no direct use of a formal notion of knowledge.
The connection with a formal epistemic model is made in \cref{sec:epist-found}.
All iterations of the considered operators start at the initial restriction $(S_1, \dots, S_n)$.

\subsection{Full communication}
\label{sec:static-setting}

Let us assume that within each hyperarc $\hyarc\in H$,
all players in $A$ have shared all information about their preferences.
We leave the precise definition of communication to \cref{sec:intermediate-states}.



For each group of players $\group \subseteq \agents$, let $S_\group$ denote the set of those restrictions of~$\G$
which only restrict the strategy sets of players from~$\group$.
That is,
\[
S_\group := \{(S'_1, \dots, S'_n) \mid S'_i \subseteq S_i \mbox{ for } i \in \group \mbox{ and } S'_i = S_i \mbox{ for } i \not\in \group\}\mpunct.
\]

We now define a contracting operator~$T_\group$ on the set~$S_\group$
as follows.
For each  $\G' = (S'_1, \dots, S'_n)\in S_\group$, let
$ 
T_\group(\G') := (S''_1, \dots, S''_n),
$ 
where for all $i \in N$,
\[
        S''_i :=
        \left\{
        \begin{array}{l@{\extracolsep{3mm}}l}
        \{s_i \in S'_i \mid \sdlocal(s_i, \G')\}     & \mathrm{if}\  i \in \group \\
        S'_i       & \mathrm{otherwise.}
        \end{array}
        \right.
\]
We call $T_\group^\infty$ the
\dfn[]{outcome of IESDS on~$\group$}.
We then define the restriction $\G(H)$ of $\G$ as
\footnote{Here and elsewhere we omit the outer brackets preceding
the outer subscript `$_i$', so $\G(H)_i$ denotes $(\G(H))_i$, etc.}
$ 
\G(H):=(\G(H)_1,\dots,\G(H)_n),
$ 
where for all $i \in N$,
\[
\G(H)_i := T_{\{i\}}\big(\textstyle\bigcap_{\hyarc: i \in \hyarc \in H} T^{\infty}_\hyarc\big)_i\mpunct.
\]
That is, the $i$th component of $\G(H)$ is the $i$th component of the result of
applying $T_{\{i\}}$ to the intersection of $T_\hyarc^\infty$
for all~$\hyarc\in H$ containing~$i$.
We call $\G(H)$ the \index{iterated elimination!outcome w.r.t.~$H$}%
\dfn[outcome!of IESDS w.r.t.~$H$]{outcome of IESDS
with respect to~$H$}.
Note that $T$ is contracting.

Let us \qinf{walk through} this definition to understand it better. Given a player $i$ and a
hyperarc $\hyarc \in H$ such that $i \in \hyarc$, $T^{\infty}_\hyarc$ is the outcome
of IESDS on $\hyarc$,
starting at $(S_1, \dots, S_n)$. The strategies
of players from outside of $\hyarc$ are not affected by this process.
This elimination process is performed simultaneously
for each hyperarc that~$i$ is a member of.
By intersecting the outcomes, i.e.,
by considering the restriction $\bigcap_{\hyarc: i \in \hyarc \in H} T^{\infty}_\hyarc$,
one arrives at a restriction in which all such \qinf{groupwise}
iterated eliminations have taken place.
However, in this restriction some of the strategies of player~$i$
may not survive elimination. They are eliminated using one application of the 
$T_{\{i\}}$ operator. We illustrate this process, and in particular this last step,
 in the following example.

\begin{figure}
  \centering
  \beginpgfgraphicnamed{ex-combining-step}%
  \begin{tikzpicture}
    \path (0,0)  node (a) {};
    \newcommand{\pica}[2]{
      \path[commgraphnode] #1 node[draw] (#2_1) {1}
        +(-3,0) node[draw] (#2_2) {2}
        +(3,0) node[draw] (#2_3) {3};
      \begin{pgfonlayer}{background}
        \node[commgraphhyperarc,transform shape,fit=(#2_1)(#2_2)] (#2_12) {};
        \node[commgraphhyperarc,transform shape,fit=(#2_1)(#2_3)] (#2_13) {};
      \end{pgfonlayer}
    }
    \pica{(a)}{a}
    \node[commonknowledge,below=of a_12,callout absolute pointer={(a_12.south)}]
         {$\preferred{}{L}{R}$};
    \node[commonknowledge,below=of a_13,callout absolute pointer={(a_13.south)}]
         {$\preferred{}{A}{B}$};
    \node[privateknowledge,above right=of a_1,callout absolute pointer={(a_1.north east)}]
         {$\preferred{}{U}{D}$};
  \end{tikzpicture}%
  \endpgfgraphicnamed
  \caption[toc entry]{Illustrating \cref{ex:combining-step}. Hyperarcs are shown in gray.
    Callouts attached to hyperarcs represent
    communicated, and thus commonly known, information.
    The thought bubble represents private information,
    in this case obtained from the combination of information only available to player~$1$.}
  \label{fig:ex:combining-step}
\end{figure}
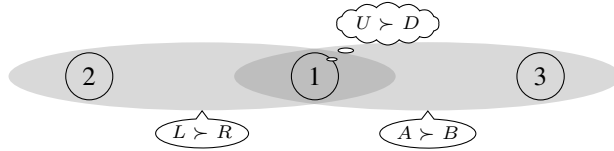

\begin{example}
  \label{ex:combining-step}
  Consider the following three-player game $\G$
  where player~$3$ chooses between the left ($A$) and the right ($B$) table.

  \begin{center}
    \hfill
    \begin{game}{2}{2}[Pl. $1$][Pl. $2$][$A$]
          \> $L$   \> $R$   \\
      $U$ \> $1,1,1$ \> $0,0,1$ \\
      $D$ \> $0,1,1$ \> $1,0,1$~
    \end{game}
    \hfill
    \begin{game}{2}{2}[][Pl. $2$][$B$]
          \> $L$   \> $R$   \\
      $U$ \> $0,1,0$ \> $0,0,0$ \\
      $D$ \> $1,1,0$ \> $1,0,0$~
    \end{game}
    \hfill~
  \end{center}
So, for example, the payoffs for the strategy profile $(U,L,B)$ are, respectively,
  $0$, $1$, and~$0$.
  Now assume the interaction structure $H=\{\{1,2\},\{1,3\}\}$.
  We obtain the outcomes $T^\infty_{\{1,2\}}=(\{U,D\},\{L\},\{A,B\})$ and $T^\infty_{\{1,3\}}=(\{U,D\},\{L,R\},\{A\})$.
  The restriction defined by these two outcomes is $(\{U,D\}, \{L\},\{A\})$,
  and in the final step player~$1$ eliminates his strategy~$D$ by one application of $T_{\{1\}}$.
  The outcome of the whole process is thus $\G(H)=(\{U\},\{L\},\{A\})$.
  See \cref{fig:ex:combining-step} for an illustration of this situation.
\end{example}

In this example, the outcome with respect to the given interaction structure
coincides with the outcome of the customary IESDS applied to the initial game.
However, this is not the case in general,
and the purpose of this example is simply to illustrate how the operators work.
\Cref{ex:h-influences-outcome} later on shows in a different setting
how the interaction structure can affect the outcome.

Note that when $H$ consists of the single hyperarc $N$ that contains
all the players, then for each player~$i$, $\bigcap_{\hyarc: i \in \hyarc \in H}
T^{\infty}_\hyarc$ reduces to $T^{\infty}_N$, and this is
closed under application of each operator $T_{\{i\}}$.  So then
$\G(H) = T^{\infty}_N$, that is, $\G(H)$ 
coincides then with the customary outcome of
iterated elimination strictly dominated strategies.

In general, this customary outcome
is included in the outcome w.r.t.~any hypergraph~$H$, as the following result shows.

\begin{theorem} \label{thm:hyper}
For any game $\G$ and hypergraph $H$ we have $T^{\infty}_N \subseteq \G(H)$.
\end{theorem}
Noting that $T^{\infty}_N=\G(\{N\})$, this can be interpreted as a special case of a monotonicity property:
If $H$ is a refinement of $H'$, then~$H'$ allows more communication to take place,
and thus more strategies to be eliminated, than~$H$. In symbols, $\G(H')\subseteq \G(H)$.
Since we are not going to use this property, we restrict ourselves to the simpler version as stated.
\begin{movedProof}
\begin{proof}[\proofof\cref{thm:hyper}]
We first prove the corresponding claim for the global version of strict dominance, \sdglobal,
introduced before \cref{lem:l-equals-g}.

Consider $\group \subseteq \group'\subseteq N$.
By definition, this implies that
for all restrictions $\G'$ we have $T_{\group'}(\G') \subseteq T_\group(\G')$.
Since $\sdglobal$ is monotonic, so is the operator $T_C$ for all $C \subseteq N$.
Hence by a straightforward induction $T_N^{\infty} \subseteq T_\group^{\infty}$ for all $\group \subseteq N$,
and consequently, for all players $i$,
\begin{equation}
\label{eqn:comminclu}
T_N^{\infty} \subseteq \textstyle\bigcap_{\hyarc: i \in \hyarc \in H} T^{\infty}_\hyarc\mpunct.
\end{equation}
Hence, for all $i \in N$,
\[
T_N^{\infty} = T_{\{i\}}(T_N^{\infty}) \subseteq T_{\{i\}}(\textstyle\bigcap_{\hyarc: i \in \hyarc \in H} T^{\infty}_\hyarc),
\]
where the inclusion holds by the monotonicity of $T_{\{i\}}$.
Consequently $T^{\infty}_N \subseteq \G(H)$.

We now prove the original claim for the local version of strict dominance.
We need to distinguish the $T_C$ operator for $\sdlocal$ and $\sdglobal$. In the former
case we write $T_{C, l}$ and in the latter case $T_{C, g}$. The reason that we use the latter
operators is that they are monotonic and closely related to the former operators.
As a consequence of \cref{lem:l-equals-g}, $T^{\infty}_{C, l} = T^{\infty}_{C, g}$.
Now fix an arbitrary $i \in N$, then
\[
\textstyle\bigcap_{\hyarc: i \in \hyarc \in H} T^{\infty}_{\hyarc,g} = \textstyle\bigcap_{\hyarc: i \in \hyarc \in H} T^{\infty}_{\hyarc,l}\mpunct,
\]
and by \eqref{eqn:comminclu} for $\sdglobal$, $T^{\infty}_{N, g} \subseteq \bigcap_{\hyarc: i \in \hyarc \in H} T^{\infty}_{\hyarc,g}$, so
\begin{equation}
  \label{equ:lg}
T^{\infty}_{N, l} = T^{\infty}_{N, g} \subseteq \textstyle\bigcap_{\hyarc: i \in \hyarc \in H} T^{\infty}_{\hyarc,l}\mpunct.
\end{equation}
Further, we have
$T^{\infty}_{N, l} = T^{\infty}_{N, g}$ and 
$T_{N, g}^{\infty} = T_{\{i\},g}(T_{N, g}^{\infty})$, so 
$T_{N, l}^{\infty} = T_{\{i\},g}(T_{N, l}^{\infty})$.
Hence, by \eqref{equ:lg} and monotonicity of $T_{\{i\},g}$,
\[
T^{\infty}_{N, l} = T_{\{i\},g}(T_{N, l}^{\infty}) \subseteq T_{\{i\},g}(\textstyle\bigcap_{\hyarc: i \in \hyarc \in H} T^{\infty}_{\hyarc,l})\mpunct.
\]
Also, for all $i \in N$ and all restrictions $\G'$ we have, by definition,
\[T_{\{i\}, g}(\G') \subseteq T_{\{i\}, l}(\G')\mpunct,\]
so by the last inclusion
\[
T^{\infty}_{N, l} \subseteq T_{\{i\},l}(\textstyle\bigcap_{\hyarc: i \in \hyarc \in H} T^{\infty}_{\hyarc,l})\mpunct.
\]
Consequently, $T^{\infty}_{N,l} \subseteq \G(H)$, as desired.

\end{proof}
\end{movedProof}



The inclusion proved in this result cannot be reversed, even when each
pair of players shares a hyperarc.  The following example also shows
that the hypergraph structure is more informative than the
corresponding graph structure.

\begin{example}
\label{ex:h-influences-outcome}
Consider the following  game with three players,
where player~$3$ chooses between the left ($A$) and the right ($B$) table.

\begin{center}
  \hfill
    \begin{game}{2}{2}[Pl. $1$][Pl. $2$][$A$]
          \> $L$   \> $R$   \\
      $U$ \> $0,1,0$ \> $0,0,1$ \\
      $D$ \> $1,0,0$ \> $1,1,1$~
    \end{game}
  \hfill
    \begin{game}{2}{2}[][Pl. $2$][$B$]
          \> $L$   \> $R$   \\
      $U$ \> $0,1,1$ \> $0,0,0$ \\
      $D$ \> $1,0,1$ \> $1,1,0$~
    \end{game}
    \hfill~
  \end{center}


  Note that the payoffs of player~$1$ and~$2$ depend only on each other's choices,
  and the payoff of player~$3$ depends only on the choices of player~$2$ and~$3$.

  If we assume the hypergraph $H$ that consists of the single hyperarc $\{1, 2, 3\}$, then
  the outcome
  of IESDS w.r.t.~$H$ is the customary outcome
  which equals $(\{D\}, \{R\}, \{A\})$. Indeed,
  player~$1$ can eliminate his strictly dominated strategy $U$, then
  player~$2$ can eliminate $L$, and subsequently player~$3$ can eliminate $B$.

  In contrast, if the hypergraph consists of all pairs of players,
  that is, $H = \{\{1, 2\},$ $\{2, 3\}, \{1, 3\}\}$, then the outcome of
  IESDS w.r.t.~$H$
  equals $(\{D\}, \{R\}, \{A,B\})$.


Indeed, in order for $3$ to be able to eliminate $B$, he would have to
know that $2$ has eliminated $L$.
However, the second hypergraph $H = \{ \{1,2\}, \{1,3\}, \{2,3\} \}$
does not allow this:
although $2$ and $3$ can both learn that $1$ eliminates $U$, they
learn it \emph{together}.
So in particular the information that $2$ knows that $1$ eliminates
$U$ is not available to $3$,
who thus does not learn that she eliminates $L$.

More formally, the result can be calculated as follows:
\begin{eqnarray*}
T_{\{1\}}( \cap_{A: i \in A \in H} T_A^\infty )_{1}
& = & T_{\{1\}} (T_{\{1,2\}}^\infty \cap T_{\{1,3\}}^\infty) _{1} \\
& = & T_{\{1\}} ((\{D\}, \{R\}, \{A, B\}) \cap (\{D\}, \{L,R\}, \{A,B\}))_{1} \\
& = & T_{\{1\}} (\{D\}, \{R\}, \{A, B\})_{1} \\
& = & \{D\}, \\
T_{\{2\}}( \cap_{A: i \in A \in H} T_A^\infty )_{2}
& = & T_{\{2\}} (T_{\{1,2\}}^\infty \cap T_{\{2,3\}}^\infty) _{2} \\
& = & T_{\{2\}} (\{D\}, \{R\}, \{A, B\}) \cap (\{U,D\}, \{L, R\}, \{A,
B\})_{2} \\
& = & \{R\}, \textup{ and } \\
T_{\{3\}}( \cap_{A: i \in A \in H} T_A^\infty )_{3}
& = & T_{\{3\}} (T_{\{1,3\}}^\infty \cap T_{\{1,3\}}^\infty) _{3} \\
& = & T_{\{3\}} (\{U,D\}, \{L,R\}, \{A, B\}) \cap ((\{D\}, \{L, R\},
\{A, B\}))_{3} \\
& = & \{A,B\}.
\end{eqnarray*}
\end{example}

\subsection{Partial communication}
\label{sec:intermediate-states}

The setting considered in \cref{sec:static-setting} corresponds to a state
in which in all hyperarcs all players have 
shared all information about their preferences.
Given the game~$\G$ and the hypergraph~$H$,
the outcome $\G(H)$ there defined thus reflects which
strategies players can eliminate if initially they know only their own preferences
and they communicate all their preferences in~$H$.
We now define formally what communication we assume possible,
and then look at states in which only certain preferences have been communicated.
Such states may be stages on the way to full communication,
but may also reflect situations in which communication has been incomplete.


Each player $i$ can communicate his preferences
to each $\hyarc\in H$ with $i\in \hyarc$.
A \dfn{message} by $i$ consists of a preference statement $\prf is'$
for $s_i,s_i'\in S_i$ and $s_{-i}\in S_{-i}$.
We denote such a message by $\msg{i}{\hyarc}{\prf is'}$
and require that $i\in\hyarc$ and that
it is \dfn{truthful} with respect to the given initial game~$\G$,
that is, indeed $\prf is'$ in $\G$.
Note that the fact that $i$ is the sender is, strictly speaking, never used.
Thus, in accordance with the interpretation of communication described in \cref{sec:motivation},
we could drop the sender and simply write ``the players in $\hyarc$ commonly observe that $\prf is'$.''
A \dfn{partial (communication) state} is now given by
the set $\msgs$ of messages which have been communicated.

We now adjust the definition of strict dominance to account for partial states,
such that in a partial state given by~$\msgs$ it uses only information shared among a given group~$\group$.
So with singleton $\group=\{i\}$ only $i$'s preferences are used,
and with larger $\group$ only preferences contained in messages to a superset of $\group$ are used:
\begin{itemize}
\item 
$\sdlocal_{\{i\},\msgs}(s_i, \G')$ holds iff $\neg\mexists s'_i\in S_i'\:\mforall s_{-i}\in S'_{-i}\:\prf is'$,

\item 
$\sdlocal_{\group,\msgs}(s_i, \G')$ holds iff $\neg\mexists s'_i\in S_i'\:\mforall s_{-i}\in S'_{-i}\:\msgs\restr_\group\vDash\prf is'$,

where $\group \neq \{i\}$ and by $\msgs\restr_\group\vDash\prf is'$ we mean that $\prf is'$
is entailed by those messages in~$\msgs$ which $\group$ received.
\end{itemize}
More precisely, the \dfn{entailment relation}
\[ 
\msgs\restr_\group\vDash\prf is'
\] 
holds iff there exist messages
$\msg{\cdot}{\group^k}{\pref{i}{s}{s^k}{s^{k+1}}}\in\msgs$
for $k \in \{1, \ldots, \ell-1\}$ such that $\group^k\supseteq \group$, $s_i^1=s_i'$ and $s_i^\ell=s_i$.

We now define a generalization of the $T_\group$ operator by: 
\[
T_{\group,\msgs}(\G') := (S''_1, \dots, S''_n),
\]
where $\G' = (S'_1, \dots, S'_n)$ and for all $i \in N$,
\[
        S''_i :=
        \{s_i \in S'_i \mid \sdlocal_{\group,\msgs}(s_i, \G')\}.
\]
Note that, as before, $S_i'$ remains unchanged for $i\not\in \group$,
since then $\sdlocal_{\group,\msgs}(s_i,\G')$ always holds.
Indeed, for it to be false, there would have to be some message $\msg{i}{\group}{\cdot}\in\msgs$,
which would imply $i\in \group$.

Similarly, we now define the
\index{iterated elimination!outcome w.r.t.~$H,\msgs$}%
\dfn[outcome!of IESDS w.r.t.~$H,\msgs$]%
{outcome of IESDS with respect to~$H,\msgs$}
to be the restriction $\G(H,\msgs)$,
where for $i \in N$
\[
\G(H,\msgs)_i := \left(T_{\{i\},\msgs}\left(\textstyle\bigcap_{\hyarc: i \in \hyarc \in\overline{H}} T^{\infty}_{\hyarc,\msgs}\right)\right)_i\mpunct.
\]
Here $\overline{H}$ denotes the closure of~$H$ under non-empty intersection.
That is, 
\[
\overline{H} = \{A_1 \cap \dots \cap A_k \mid \{A_1, \dots, A_k\} \subseteq H\} \setminus \{\emptyset\}.
\]

The use of $\overline{H}$ is necessary because
certain information may be entailed by messages sent to different
hyperarcs.  For example, with
$\msg{j}{A}{\pref{j}{s}{s''}{s'}},\msg{j}{A'}{\prf js'}\in\msgs$, the
combined information that $\prf{j}{s}{''}$ is available to the members of $A\cap A'$.


Again, let us \qinf{walk through} the definition of $\G(H,\msgs)$.
First, a separate elimination process is run on each hyperarc of $\overline{H}$, using only
information which has been communicated there (which now no longer covers all
members' preferences, but only the ones according to the partial state $\msgs$).
Then, in the final step, each player combines his insights from all hyperarcs
of which he is a member, and eliminates any strategies
that he thereby learns not to be optimal.

Note that the underlying hypergraph $H$ can be \qinf{recovered} from the
set of messages $M$ used, that is, we could define $H := \{A \mid (.,
A, .) \in M\}$.  However, explicit use of the hypergraph allows us to
compare the setting with partial information with that of full
information. Namely, it is easy to see that in the case where the
players have communicated all there is to communicate, i.e., for
\[
\allmsgs{H}:=\{\msg{i}{\hyarc}{\prf is'} \mid i\in\agents, \hyarc\in H,\text{ $s_i,s_i'\in S_i$ with $\prf is'$ in $\G$}\}\mpunct,
\]
the partial outcome coincides with the previously defined outcome, i.e.,
\[
\G(H,\allmsgs{H})=\G(H).
\]
This corresponds to the intuition that $\G(H)$ captures the elimination process
when all possible communication has taken place.
In particular, all entailed information has also been communicated in $\allmsgs{H}$,
which is why we did not need to consider $\overline{H}$ in \cref{sec:static-setting}.


The effect of full communication within a hyperarc $A$ is that the
players in $A$ acquire common knowledge of the set $T^{\infty}_A$.
However, to compute the final outcome of the elimination process, that is
$\G(H)$,  each player combines this information across all hyperarcs
he is a member of. Consequently, the players do not have common
knowledge of $\G(H)$. This point is illustrated in the following
example.

\begin{figure}
  \centering
  \beginpgfgraphicnamed{ex-intermediate-states}%
  \begin{tikzpicture}[rotate=90]
    \path (0,0)  node (a) {} ++(0,-3) node (b) {} ++(0,-3.2) node (c) {} ++(0,-3.2) node (d) {};
    \newcommand{\pica}[2]{
      \path[commgraphnode] #1 node[draw] (#2_1) {1}
        +(1.5,0) node[draw] (#2_2) {2}
        +(-1.5,0) node[draw] (#2_3) {3};
      \begin{pgfonlayer}{background}
        \node[commgraphhyperarc,transform shape,fit=(#2_1)(#2_2)] (#2_12) {};
        \node[commgraphhyperarc,transform shape,fit=(#2_1)(#2_3)] (#2_13) {};
      \end{pgfonlayer}
    }
    \pica{(a)}{a}
    \node[privateknowledge,above right=of a_2,callout absolute pointer={(a_2.north east)}]
         {$\preferred{s_{-2}}{L}{R}$};
    \node[privateknowledge,above right=of a_3,callout absolute pointer={(a_3.north east)}]
         {$\preferred{s_{-3}}{A}{B}$};
    \newcommand{\picb}[2]{
      \pica{#1}{#2}
      \node[commonknowledge,right=of #2_12.320,callout absolute pointer={(#2_12.320)}]
           {$\preferred{s_{-2}}{L}{R}$};
    }
    \picb{(b)}{b}
    \node[privateknowledge,above right=of b_3,callout absolute pointer={(b_3.north east)}]
         {$\preferred{s_{-3}}{A}{B}$};
    \newcommand{\picc}[2]{
      \picb{#1}{#2}
      \node[commonknowledge,right=of #2_13.220,callout absolute pointer={(#2_13.220)}]
           {$\preferred{s_{-3}}{A}{B}$};
    }
    \picc{(c)}{c}
    \newcommand{\picd}[2]{
      \picc{#1}{#2}
      \node[privateknowledge,above right=of #2_1.east,callout absolute pointer={(#2_1.20)}]
           {$\preferred{}{U}{D}$};
    }
    \picd{(d)}{d}
    \foreach \n/\nn/\l in {a/b/(a),b/c/(b),c/d/(c)}
      \path (\n_13.west) edge[nextpic]
        (\nn_13.west);
    \node[below of=a_3] {$\msgs$};
    \node[below of=b_3] {$\msgs'$};
    \node[below of=c_3] {$\msgs''$};
  \end{tikzpicture}%
  \endpgfgraphicnamed    
  \caption[toc entry]{Illustrating \cref{ex:intermediate-states}.}
  \label{fig:ex:intermediate-states}
\end{figure}
\begin{example}
  \label{ex:intermediate-states}
  The process described in this example is illustrated in \cref{fig:ex:intermediate-states}.
  Consider again the game $\G$ from \cref{ex:combining-step},
  and the initial state where $\msgs=\emptyset$.
  We have $T^\infty_{\hyarc,\msgs}=\G$ for all $\hyarc\in \overline{H}$,
  that is, without communication
  no strategy can \qinf{commonly} be eliminated. 
  However, players~$2$ and~$3$ can \qinf{privately} eliminate one of their strategies each,
  since each of them knows his own preferences, so
  \begin{align*}
    T_{\{1\},\msgs}(\textstyle\bigcap_{\hyarc: 1 \in \hyarc \in \overline{H}} T^{\infty}_{\hyarc,\msgs})&=(\{U,D\},\{L,R\},\{A,B\})\mpunct,\\
    T_{\{2\},\msgs}(\textstyle\bigcap_{\hyarc: 2 \in \hyarc \in \overline{H}} T^{\infty}_{\hyarc,\msgs})&=(\{U,D\},\{L\},\{A,B\})\mpunct,\\
    T_{\{3\},\msgs}(\textstyle\bigcap_{\hyarc: 3 \in \hyarc \in \overline{H}} T^{\infty}_{\hyarc,\msgs})&=(\{U,D\},\{L,R\},\{A\})\mpunct,\\
  \end{align*}
  This elimination cannot be iterated further by other players and the overall outcome is
$ 
    \G(H,\msgs)=(\{U,D\},\{L\},\{A\})\mpunct.
$ 

  Consider now the partial state
$ 
    \msgs'=\{\msg{2}{\{1,2\}}{\preferred{s_{-2}}{L}{R}}\suchthat s_{-2}\in S_{-2}\}\mpunct,
$ 
  that is, a state in which player~$2$ has shared with player~$1$ the information
  that for any joint strategy of players~$1$ and~$3$,
  he prefers his strategy $L$ over $R$.
  Then only the result of player~$1$ changes:
  \begin{align*}
    T_{\{1\},\msgs'}(\textstyle\bigcap_{\hyarc: 1 \in \hyarc \in \overline{H}} T^{\infty}_{\hyarc,\msgs'})&=(\{U,D\},\{L\},\{A,B\})\mpunct,
  \end{align*}
  while the other results and the overall outcome remain the same.
  If additionally player~$3$ communicates all his information in the hyperarc he shares with player~$1$,
  that is, if the partial state is
$ 
    \msgs''=\msgs'\cup\{\msg{3}{\{1,3\}}{\preferred{s_{-3}}{A}{B}}\suchthat s_{-3}\in S_{-3}\}\mpunct,
$ 
  then player~$1$ can combine all the received information and obtain
  \begin{align*}
    T_{\{1\},\msgs''}(\textstyle\bigcap_{\hyarc: 1 \in \hyarc \in \overline{H}} T^{\infty}_{\hyarc,\msgs''})&=(\{U\},\{L\},\{A\})\mpunct.
  \end{align*}
  This is also the overall outcome $\G(H,\msgs'')$,
  coinciding with the outcome $\G(H,\allmsgs{H})$
  where all possible information has been communicated.
\end{example}

Let us now illustrate the importance of using entailment in the optimality notions for partial states
and $\overline{H}$ (rather than~$H$) in the definition of $\G(H,\msgs)$.
\begin{figure}[t]
  \centering
  \beginpgfgraphicnamed{ex-hbar-matters}%
    \begin{tikzpicture}[rotate=90,scale=1]
    \path (0,0)  node (a) {} ++(0,-2.8) node (b) {} ++(0,-3.1) node (c) {} ++(0,-2.9) node (d) {};
    \newcommand{\pica}[2]{
      \path[commgraphnode] #1 node[draw] (#2_1) {1}
        +(-1.5,0) node[draw] (#2_3) {3}
        ++(1.5,0) node[draw] (#2_2) {2}
        +(1.5,0) node[draw] (#2_4) {4};
      \begin{pgfonlayer}{background}
        \path (#2_3) -- node[commgraphhyperarc,sloped,transform shape,inner sep=1pt,fit=(#2_1)(#2_2.center)(#2_3.center)] (#2_123) {} (#2_2);
        \path (#2_4) -- node[commgraphhyperarc,sloped,transform shape,inner sep=1pt,fit=(#2_1.center)(#2_2)(#2_4.center)] (#2_124) {} (#2_1);
      \end{pgfonlayer}
    }
    \pica{(a)}{a}
    \setbox0=\hbox{$\preferred{R}{A}{C}$}
    \node[commonknowledge,right=of a_123.200,callout absolute pointer={(a_123.200)},text width=\wd0,text centered]
         {$\preferred{L}{A}{B}$\\$\preferred{R}{A}{C}$};
    \node[commonknowledge,right=of a_124.340,callout absolute pointer={(a_124.340)}]
         {$\preferred{L}{B}{C}$};
    \newcommand{\picb}[2]{
      \pica{#1}{#2}
      \node[commgraphhyperarc,draw,opacity=1,inner sep=2pt,thick,dashed,fill=none,transform shape,fit=(#2_1)(#2_2)] (#2_12) {};
    }
    \picb{(b)}{b}
    \node[commonknowledge,right=of b_12.south,callout absolute pointer={(b_12.south)}]
         {$\leadsto\preferred{}{A}{C}$};
    \newcommand{\picc}[2]{
      \picb{#1}{#2}
      \node[commonknowledge,right=of #2_12.south,callout absolute pointer={(#2_12.south)}]
           {$\preferred{}{A}{C}$};
      \node[commonknowledge,right=of #2_123.200,callout absolute pointer={(#2_123.200)}]
           {$\preferred{-C}{R}{L}$};
    }
    \picc{(c)}{c}
    \newcommand{\picd}[2]{
      \picc{#1}{#2}
      \node[privateknowledge,right=of #2_1,xshift=-5pt,yshift=-5pt,callout absolute pointer={(#2_1.north east)}]
           {$\preferred{}{D}{A,B,C}$};
    }
    \picd{(d)}{d}
    \foreach \n/\nn/\l in {a/b/(a),b/c/(b),c/d/(c)}
      \path (\n_123.west) edge[nextpic]
      (\nn_123.west);
    \node[below of=a_3] {$\msgs$};
    \node[below of=c_3] {$\msgs'$};
  \end{tikzpicture}%
  \endpgfgraphicnamed
  \caption[toc entry]{Illustrating \cref{ex:hbar-matters}.
    Strategies of the dummy players are omitted.
    $\preferred{}{A}{C}$ stands for $\preferred{s_{-1}}{A}{C}$,
    and $\preferred{-C}{}{}$ combines $\preferred{\alpha}{}{}$ for $\alpha\in\{A,B,D\}$.
    Note that in the first step, 
    information is not explicitly communicated but deduced.}
  \label{fig:ex:hbar-matters}  
\end{figure}
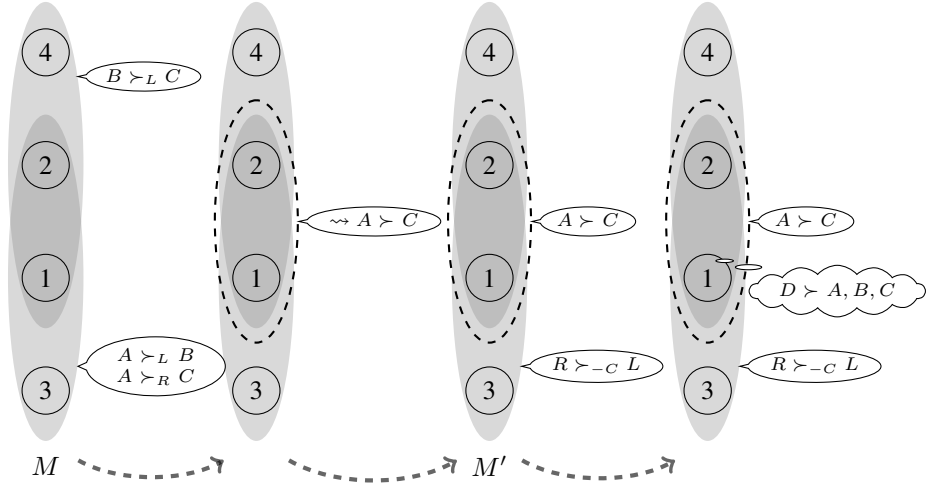
\begin{example}
\label{ex:hbar-matters}
  We look at a game involving four players,
  but we are only interested in the preferences of two of them.
  The other two players serve merely to create different hyperarcs.
  The strategies and payoffs of player~$1$ and~$2$ are as follows:

  \begin{center}
    \begin{game}{4}{2}[Pl. $1$][Pl. $2$]
          \> $L$   \> $R$   \\
      $A$ \> $3,0$ \> $1,1$ \\
      $B$ \> $2,0$ \> $1,1$ \\
      $C$ \> $1,1$ \> $0,0$ \\
      $D$ \> $0,0$ \> $5,1$~
    \end{game}
  \end{center}\vspace{5mm}

  For players~$3$ and~$4$ we assume a \qinf{dummy} strategy, denoted respectively by~$X$ and~$Y$.
  Consider the hypergraph
$ 
  H=\{\{1,2,3\},\{1,2,4\}\}
$ 
  and the partial state
  \begin{align*}
    \msgs=\{&\msg{1}{\{1,2,3\}}{\preferred{LXY}{A}{B}},\\
    &\msg{1}{\{1,2,4\}}{\preferred{LXY}{B}{C}},\\
    &\msg{1}{\{1,2,3\}}{\preferred{RXY}{A}{C}}\}\mpunct.
  \end{align*}
  The fact that player~$1$, independently of what the remaining players do,
  strictly prefers~$A$ over~$C$ is not explicit in these pieces of information,
  but it is \emph{entailed} by them,
  since $\preferred{LXY}{A}{B}$ and $\preferred{LXY}{B}{C}$ imply $\preferred{LXY}{A}{C}$.
  However, this combination of information is only available to the players in~$\{1,2,3\}\cap\{1,2,4\}$.

  Player~$2$ can make use of this fact that~$C$ is dominated,
  and eliminate his own strategy~$L$.
  If we now look at a state in which player~$2$ has communicated his relevant preferences, so
$ 
  \msgs'=\msgs\cup\{\msg{2}{\{1,2,3\}}{\preferred{\alpha XY}{R}{L}}\suchthat\alpha\in\{A,B,D\}\}\mpunct,
$ 
  we notice that player~$1$ can in turn eliminate~$A$ and~$B$,
  but only by combining information available to the players in~$\{1,2,3\}\cap\{1,2,4\}$.
  There is no single hyperarc in the original hypergraph which has all the required information available.
  It thus becomes clear that we need to take into account iterated elimination on intersections of hyperarcs.

  The whole process is illustrated in \cref{fig:ex:hbar-matters}.
\end{example}

\section{Epistemic foundations}
\label{sec:epist-found}
In this section, we provide epistemic foundations for our framework.
The aim is to prove that the definition of the outcome $\G(H,\msgs)$ correctly captures
what strategies the players can eliminate using all they \qinf{know}, in a formal sense.

We proceed as follows.  First, in \cref{sec:epistemic-model}, we
briefly introduce an epistemic model formalizing the players'
knowledge.   We apply the basic framework and results from
\cite{apt_knowledge_2009}.
In \cref{sec:epist-form-strict}, we give a general
epistemic formulation of strict dominance and argue that it correctly
captures the notion.  \Cref{sec:epist-form-strict} also contains the
main result of our epistemic analysis, namely that the outcome
$\G(H,\msgs)$ indeed yields the outcome stipulated by the epistemic
formulation.

\subsection{Epistemic language and states}
\label{sec:epistemic-model}

Again, we assume a fixed game $\G$ with non-empty set of
strategies $S_i$ for each player~$i$,
and a hypergraph~$H$ representing the interaction structure.
Analogously to \cite{apt_knowledge_2009},
we use a propositional epistemic language
with a set $\atoms$ of \dfn[atomic!proposition]{atoms} which
is divided into disjoint subsets $\atoms_i$, one for each player $i$,
where
$ 
\atoms_i := \{\prf is' \mid s_i,s_i'\in S_i, s_{-i} \in S_{-i}\}.
$ 

The set $\atoms_i$ describes all possible strict preferences between pairs of strategies of player
$i$, relative to a joint strategy of the opponents.
A \dfn{valuation} $\init$ is a subset of the atoms $\atoms$ such that for each
$s_{-i}\in S_{-i}$, the restriction
$\init\cap\{\preferred{s_{-i}}{\cdot}{\cdot}\}$ is a strict
partial order.
\footnote{
In \cite{apt_knowledge_2009} we did not consider such restrictions on valuations;
however, the relevant results can easily be seen to remain correct.
See also~\cite[Chapter 2]{witzel_knowledge_2009}.
}

Intuitively, a valuation consists of exactly the atoms assumed true.
Each specific
game $\G$ \dfnless[valuation!induced by game]{induces} exactly one
valuation which simply represents its preferences. However, in general
we also need to model the fact that players may not have full
knowledge of the game.  The restriction imposed on the valuations 
ensures that each of them is induced by some game.

So for example $\{\preferred{a}{s}{t}\}$ is a valuation
(given a game with appropriate strategy sets),
while $\{\preferred ast,\preferred atu\}$ and $\{\preferred ast,\preferred ats\}$ are not
(the former failing to make $\preferred asu$ true).

Recall from \cref{sec:intermediate-states} that a \dfn{message}
from player~$i$ to a hyperarc $A\in H$ has the form
$\msg{i}{\hyarc}{\prf is'}$, where $i\in \hyarc$, $s_i,s_i'\in S_i$, and $s_{-i}\in S_{-i}$.
We say that a message $\msg{\cdot}{\cdot}{p}$ is \dfn{truthful}
with respect to a valuation~$\init$
if $p\in\init$.
A \dfn[state!of the world]{state}, or \dfn{possible world}, is a pair
\state, where $\init$ is a valuation and \msgs is a set of messages
that are truthful with respect to~$\init$.


For a set of messages $\msgs$, $\group\subseteq N$, and $p\in\atoms$,
$\msgs\restr_\group\vDash p$ is defined as in \cref{sec:intermediate-states}.
That is, $\msgs\restr_\group\vDash p$
means that $p$ is entailed by the messages in $\msgs$ received by $\group$,
for example, by transitivity of the represented preference order.

In~\cite{apt_knowledge_2009}, we defined set operations to act component-wise on states, e.g.
$\state \subseteq \state[']$ iff $\init \subseteq \init'$ and $\msgs \subseteq \msgs'$.
However, the results we consider also hold with a modified inclusion relation,
where $\msgs\subseteq\msgs'$ iff
for each $\msg{i}{\hyarc}{p}\in\msgs$ there is $\msg{i}{\hyarc'}{p}\in\msgs'$ with $\hyarc\subseteq \hyarc'$.
By $\init_i$ we denote $\init \cap \bits_i$, the restriction of $\init$ to player $i$ atoms.
Further, $\msgs_i := \{(\cdot,A,\cdot) \in \msgs \mid i \in A\}$ is
the set of messages that player $i$ received.

We now define an \dfn{indistinguishability relation} between states:
\[
\mbox{$\state  \sim_i \state[']$ iff $(\init_i, \msgs_i) = (\init'_i, \msgs'_i)$.}
\]
For $\group \subseteq N$ the relation $\sim_\group$ is the transitive
closure of $\bigcup_{i \in \group}\sim_i$.

We consider the following positive epistemic \dfn{language} $\mathcal{L}^+$:
\[
 \varphi ::= p \mid \varphi \land \varphi \mid \varphi \lor \varphi \mid \ck \group \varphi,
\]
where the atoms $p$ denote the facts in~$\bits$,
while $\land$ and $\lor$ are the standard connectives;%
\footnote{Since we want to eliminate strategies that \emph{are} dominated,
we only need to express facts of the form $s\succ t$ (or $t\succ s$),
rather than their negated counterparts that would pertain to strategies \emph{not} being dominated.
This observation generalizes to the formulas we consider.}
and $\ck \group$ is a knowledge operator, with $\ck \group \varphi$ meaning $\varphi$
is \dfnless[]{common knowledge} among~$\group$.
The intuition for $\ck\group$ is that everybody in~$\group$ knows~$\varphi$,
everybody knows that everybody knows~$\varphi$, etc.
We write $\knows i$ for $\ck{\{i\}}$;
$\knows i \varphi$ can be read `$i$ knows that $\varphi$'.
For a sequence of players $w=i_1\ldots i_k$,
we write $\knows w$ to abbreviate $\knows{i_1}\knows{i_2} \ldots \knows{i_k}$.
Note that this language does not contain negation, which is not needed in our presentation.
We assume that each player~$i$ initially knows the true facts in $\atoms_i$
entailed by the initial game $\G$ and that the basic assumptions from \cref{sec:motivation}
are commonly known among the players.

The formal \dfn{semantics} is defined as follows
(recall that each message in \msgs is truthful with respect to~$\init$):

\begin{definition}
  \label{dfn:semantics}
  \label{result:i-knows-own}
  \begin{align*}
    \state  &\vDash p                   && \iff p \in \init, \\
    \state  &\vDash \varphi \lor \psi       && \iff \state  \vDash \varphi\text{ or }\state  \vDash \psi, \\
    \state  &\vDash \varphi \land \psi      && \iff \state  \vDash \varphi\text{ and }\state  \vDash \psi, \\
    \state &\vDash \ck \group \varphi && \iff \state['] \vDash \varphi~
    \begin{aligned}[t]
      &\text{for each \state[']}\\
      &\text{with } \state \sim_{\group} \state['].
    \end{aligned}
  \end{align*}
\end{definition}

Now we are ready to state the following results from~\cite{apt_knowledge_2009},
slightly adapted to fit our notation.


\begin{lemma}[{from \cite[Lemma 3.2]{apt_knowledge_2009}}]
  \label{result:positive-keep-holding}
  For any $\varphi\in\mathcal{L}^+$ and states \state and \state[']
  with $\state\subseteq\state[']$,
\[
    \textup{if } \state\vDash\varphi, \textup{ then } \state['] \vDash\varphi.
\]
\end{lemma}

\begin{theorem}[{from \cite[Theorem 3.5]{apt_knowledge_2009}}]
  \label{result:k-disjunction-distributes}
  \label{result:ck-disjunction-distributes}
  For any $\varphi_1,\varphi_2\in\mathcal{L}^+$, state \state, and $\group\subseteq N$,
\[
\mbox{$\state\vDash \ck \group(\varphi_1\vee\varphi_2)$ iff $\state\vDash \ck \group\varphi_1\vee \ck \group\varphi_2$.}
\]
\end{theorem}

This theorem states that in this setting, knowledge distributes over
disjunction, \emph{when the formulas in question are positive}.  That
might seem surprising, and so it is worth emphasizing that the result
does depend on the positivity of the formulas ($\varphi_1, \varphi_2
\in \mathcal{L}^+$), and also on the assumptions about communication,
and therefore possible configurations of knowledge, that are inherent
in our setup.

\begin{lemma}[{from \cite[Lemma 2.3.8]{witzel_knowledge_2009}, cf.~\cite[Lemma 3.7]{apt_knowledge_2009}}]
  \label{result:knowledge-chain-fact-equiv-msgs-equiv-ck}
  For any $\group\subseteq N$ with $\abs{\group}\geq2$,
  $p\in\bits$, 
  and state \state,
  the following are equivalent:
  \begin{enumerate}[(i)]
  \item\label{result:knowledge-chain-fact-equiv-msgs-equiv-ck:msgs}
    $\msgs\restr_\group\vDash p$,
  \item\label{result:knowledge-chain-fact-equiv-msgs-equiv-ck:ck}
    $\state\vDash \ck \group p$ 
  \end{enumerate}
\end{lemma}

\begin{theorem}[{from \cite[Theorem 3.8]{apt_knowledge_2009}}]
  \label{result:knowledge-chain-phi-equiv-ck}
  \label{thm:permutation}
  For any $\group\subseteq N$, $\varphi\in\mathcal{L}^+$, and state \state,
  \begin{center}
    $\state\vDash \ck \group\varphi$ iff $\state\vDash \knows w\varphi$ for some permutation $w$ of $\group$.
  \end{center}
\end{theorem}

\subsection{Correctness result}
\label{sec:epist-form-strict}
We use results from \cite{apt_knowledge_2009}, summed up in \cref{sec:epistemic-model},
in order to prove that the $T_G$ operator defined in \cref{sec:iter-strat-elim} is correct
with respect to an epistemic formulation of our setting.

In order to simplify our presentation, we use a \emph{global} version of strict dominance (introduced by \citet{chen_iterated_2007}) instead of \sdlocal.
We use
\[
\sdglobal(s_i, \G')\text{ to abbreviate }\neg \mexists s'_i
  \in S_i \: \mforall s_{-i}'\in S_{-i}' \: s'_i \succ_{s_{-i}'} s_i.
\]
The difference is that \sdglobal considers dominance by a strategy from the \emph{original} set $S_i$,
while \sdlocal considers dominance only by strategies from the set $S_i'$ that has survived elimination in the subgame $\G'$.
So in $sd^{g}$, 
it is stipulated that a strategy is not strictly dominated by a 
strategy \emph{from the initial game}.
As noted in~\cite{BFK06,apt_many_2007}, \sdglobal is monotonic,
while $\sdlocal$ is not.

In order to use \sdglobal instead of \sdlocal, we need to show that our definitions and results do not depend
on the choice of strict dominance relation.

We first relate the global and the local version of strict dominance,
slightly overloading notation by letting $\sdlocal(\G')=(\{s_i\in S_i' \suchthat \sdlocal(s_i,\G')\})_{i \in N}$,
and similarly for $\sdglobal$.
The two notions are equivalent in the following sense.


\begin{lemma}\label{lem:l-equals-g}
  Let $\G^0,\G^1,\dots$ be a sequence of restrictions of the initial restriction $(S_1, \dots, S_n)$,
  such that $\G^0=(S_1,\dots,S_n)$ and $\sdglobal(\G^k)\subseteq \G^{k+1}\subseteq \G^k$
  for all $k\geq0$.
  Then for all $k\geq0$, $\sdglobal(\G^k)=\sdlocal(\G^k)$.
\end{lemma}

Intuitively, this claim can be stated as follows. Suppose that each
restriction $\G^{k+1}$ is obtained from $\G^{k}$ by removing some
strategies that are strictly dominated in the global sense. Then the
local and global strict dominance coincide on each considered
restriction $\G^{k}$. As a consequence, the result also holds if the strategies removed
are required to be strictly dominated in the local instead of the global sense.

\begin{movedProof}
\begin{proof}[\proofof\cref{lem:l-equals-g}]
  We show that for all $k\geq0$ each globally (strictly) dominated
  strategy is also locally dominated in $\G^k$.  Together with the
  straightforward fact that local dominance implies global dominance,
  this proves the desired equivalence. 
  
  Formally, the claim is thus
  that for all $k\geq0$, $s_i\in S_i$ and $s_i'\in \G^k_i$ such that
  $\pref{i}{\G^k}{s}{s'}$, there is $s_i''\in \G^k_i$ with
  $\pref{i}{\G^k}{s''}{s'}$.
  To show this we prove by induction that for all $k\geq0$ and $s_i\in
  S_i$, there is $s_i''\in \G^k_i$ such that
  $\prefeq{i}{\G^k}{s''}{s}$, from which the claim follows since
$\prefeq{i}{\G^k}{s''}{s}$ and $\pref{i}{\G^k}{s}{s'}$ imply $\pref{i}{\G^k}{s''}{s'}$.

This claim clearly holds for $k = 0$.  Now assume the statement holds
for some $k$ and fix $s_i\in S_i$.  Choose some $s_i'\in S_i$ that
is $\preferredeq{\G^k_{-i}}{}{}$-maximal among the elements of $S_i$.
By the induction hypothesis there is $s_i''\in \G^k_i$ such that
$\prefeq{i}{\G^k}{s''}{s'}$.
So also $s_i''$ is $\preferredeq{\G^k_{-i}}{}{}$-maximal among the elements of $S_i$.
Hence $s_i'' \in (\sdglobal(\G^k))_i$.

  From $\sdglobal(\G^k)\subseteq \G^{k+1}$ we now obtain $s_i''\in \G^{k+1}_i$.
  From $\G^{k+1}\subseteq \G^k$ and $\prefeq{i}{\G^k}{s''}{s'}$, we obtain $\prefeq{i}{\G^{k+1}}{s''}{s'}$,
  and by the maximality of $s_i'$, we also have $\prefeq{i}{\G^{k+1}}{s'}{s}$.
  Thus, $s_i''\in \G^{k+1}_i$ and $\prefeq{i}{\G^{k+1}}{s''}{s}$, which concludes the proof of the induction step.
\end{proof}
\end{movedProof}







Next, we show that the operators we defined in \cref{sec:iter-strat-elim} produce sequences that satisfy the conditions of \cref{lem:l-equals-g},
and thus coincide for the global and the local version of strict dominance.

\begin{lemma} \label{thm:lg}
  For any hypergraph $H$ and set of messages $\msgs$,
  the outcome $\G(H, \msgs)$ does not depend on the choice of \sdlocal or \sdglobal.
\end{lemma}

\begin{movedProof}
In order to prove \cref{thm:lg}, we need an auxiliary lemma dealing with operators in a general setting.
Given $Y \in D$ and an operator $T$  on a finite lattice $(D, \subseteq)$,
we denote by $T_Y$ the following operator:
\[
T_Y(X) := T(X) \cup (X \cap Y).
\]

\begin{numberednote}\label{not:T}
  If the $T$ operator is contracting, then so is $T_Y$.
\end{numberednote}

\begin{lemma}\label{lem:TU}
  Suppose that $T$ and $U$ are operators on a finite lattice $(D,
  \subseteq)$ such that $T$ is monotonic and contracting.  Then $
  T_{T^{\infty} \cap U^{\infty}}^{\infty} (U^{\infty}) = T^{\infty}
  \cap U^{\infty} $.
\end{lemma}

Informally, this claim states that the combined effect of independent limit
iterations of $T$ and $U$ can be modelled by `serial' limit iterations of
$T$ and $U$, provided the operator $T$ is modified to an appropriate
$T_Y$ form.  
\medskip

\begin{proof}

Denote for brevity $T^{\infty} \cap U^{\infty}$ by $Y$.
First we prove by induction that for all $k \geq 0$ 
\[
Y \subseteq T_{Y}^{k} (U^{\infty}).
\]
The claim clearly holds for $k = 0$. 
Suppose it holds for some $k \geq 0$. Then by the induction hypothesis
\[
T_{Y}^{k+1}(U^{\infty}) = T(T_{Y}^{k}(U^{\infty})) \cup (T_{Y}^{k}(U^{\infty}) \cap Y) \supseteq Y.
\]
Hence
\begin{equation}
  \label{equ:TU}
Y \subseteq T_{Y}^{\infty} (U^{\infty}).
\end{equation}

To prove the converse implication we show by induction that for all $k \geq 0$ 
\[
T_{Y}^{k}(U^{\infty}) \subseteq T^{k}.
\]
The claim clearly holds for $k = 0$. 
Suppose it holds for some $k \geq 0$. Then by the induction hypothesis
and the monotonicity of $T$
\[
T_{Y}^{k+1}(U^{\infty}) = T_Y(T_{Y}^{k}(U^{\infty})) \subseteq T_Y(T^{k}) = 
T^{k+1} \cup (T^{k} \cap Y) \subseteq T^{k+1} \cup T^{\infty}  \subseteq T^{k+1}.
\]
Hence
\begin{equation}
  \label{equ:T}
T_{Y}^{\infty} (U^{\infty}) \subseteq T^{\infty}.  
\end{equation}

Next, by \cref{not:T} the operator $T_{Y}$ is contracting, so
\begin{equation}
  \label{equ:U}
T_{Y}^{\infty} (U^{\infty}) \subseteq U^{\infty}.  
\end{equation}

Now the claim follows by \eqref{equ:TU}, \eqref{equ:T} and \eqref{equ:U}. 
\end{proof}

\begin{proof}[\proofof\cref{thm:lg}]

Recall that $\overline{H}$ denotes the closure of~$H$ under non-empty intersection.
Fix an interaction structure $H$, a set of messages $\msgs$ and $i \in
N$.  Assume for simplicity that the set $\{A \mid i
\in A \in \overline{H}\}$ has exactly two elements, say, $B_i$ and
$C_i$.  To deal with the arbitrary situation \cref{lem:TU} needs
to be generalized to an arbitrary number of operators. Such a
generalization is straightforward and omitted.

Let now
\[
\begin{array}{l}
\G_1 := (S_1, \ldots, S_n),  \\
\G_2 := T^{\infty}_{B_i, \msgs}(\G_1), \\
\G_3 := \hat{T}^{\infty}_{C_i, \msgs}(\G_2), \\
\G_4 := T_{\{i\}}(\G_3), 
\end{array}
\]
where
\[
\hat{T}_{C_i, \msgs}(\G) := T_{C_i, \msgs}(\G) \cup (\G \cap T^{\infty}_{B_i, \msgs} \cap T^{\infty}_{C_i, \msgs}).
\]

Now, recall that 
 \[
 \G(H,\msgs)_i := \left(T_{\{i\},\msgs}\left(\textstyle\bigcap_{\hyarc: i \in \hyarc \in\overline{H}} T^{\infty}_{\hyarc,\msgs}\right)\right)_i\mpunct.
 \]
By \cref{lem:TU}, $\G_3 = T^{\infty}_{B_i, \msgs} \cap
T^{\infty}_{C_i, \msgs}$, so $(\G_4)_i$ is $\G(H,\msgs)_i$, the $i$th component of $\G(H,\msgs)$.
\Cref{not:T} ensures that each of the operators
$T_{C_i, \msgs}, T_{B_i, \msgs}$ and $\hat{T}_{C_i, \msgs}$
is contracting. Moreover, $\sdglobal$ removes (weakly) more strategies than each of them, so
the sequence of restrictions
\[
\G_1, \ T_{B_i, \msgs}(\G_1), \ T^{2}_{B_i, \msgs}(\G_1), \ldots, \G_2, \ 
\hat{T}_{C_i, \msgs}(\G_2), \ \hat{T}^{2}_{C_i, \msgs}(\G_2), \dots, \G_3, \ 
\G_4
\] 
satisfies the conditions of \cref{lem:l-equals-g}.
By \cref{lem:l-equals-g} we also obtain the same restriction $\G_3$ when in the
definition of the $T_{B_i, \msgs}$ and $T_{C_i, \msgs}$ operators we use
$\sdlocal$ instead of $\sdglobal$.  So the $i$th component 
$\G(H,\msgs)_i$
of 
$\G(H,\msgs)$
is the
same when in the definitions of the $T_{B_i, \msgs}$, $T_{C_i, \msgs}$ and
$T_{\{i\}}$ operators we use $sd^{\ell}$ instead of $sd^g$.

This concludes the proof.
\end{proof}
\end{movedProof}

\medskip

We now start the main task of this section by giving a formula describing the global version of
iterated elimination of strictly dominated strategies in the customary
 games.
We define, for $i\in N$ and $s_i\in S_i$,
\begin{align*}
  \domin^1(s_i) &:= \textstyle
  \bigvee_{s_i'\in S_i}\bigwedge_{s_{-i}\in S_{-i}}\prf is', \\
  \domin^{\ell+1}(s_i) &:= \textstyle
  \bigvee_{s_i'\in S_i}\bigwedge_{s_{-i}\in S_{-i}}
  \big(\prf is' \vee \bigvee_{j\in N\setminus\{i\}}\domin^\ell(s_j) \big). \\
\end{align*}

The following simple result relates this formula to the $T_N$ operator
(that is, $T_G$ where $G$ is the group of all players),
where we assume $sd^{g}$ as the optimality notion.

\begin{proposition}
For any game $\G$, $\ell \geq 1$, and $i\in N$
\[
(T_{N}^{\ell})_i = \{s_i \mid \domin^{\ell}(s_i) \text{ does not hold}\}.
\]
Consequently
\[
(T_{N}^{\infty})_i = \{s_i \mid \domin^{\infty}(s_i) \text{ does not hold}\}.
\]
(The propositional formulas here are evaluated with respect to the valuation induced by~$\G$.)
\end{proposition}

We now modify the above formula to an epistemic formula describing the
iterated elimination of strictly dominated strategies (in the sense of
$sd^{g}$) in  games with interaction structures. 
In contrast to the above formulation 
the formula below states that player $i$ \emph{knows} that a strategy
is strictly dominated.

We define, for $i\in N$ and $s_i\in S_i$,
\begin{align*}
  \dominated^1(s_i) &:= \textstyle
  \knows i\bigvee_{s_i'\in S_i}\bigwedge_{s_{-i}\in S_{-i}}\prf is', \\
  \dominated^{\ell+1}(s_i) &:= \textstyle
  \knows i\bigvee_{s_i'\in S_i}\bigwedge_{s_{-i}\in S_{-i}}
  \big(\prf is' \vee \bigvee_{j\in N\setminus\{i\}}\dominated^\ell(s_j) \big). 
\end{align*}

Note that $s_j$ refers to $j$'s component of $s_{-i}$.
Note also that $\dominated^1(s_i) = \knows i \domin^1(s_i)$ and that 
$\dominated^{\ell +1}(s_i)$ is defined in terms of
$\dominated^{\ell}(s_i)$ and not $\domin^{\ell}(s_i)$.

So, in the base case, player~$i$ knows that $s_i$ is strictly dominated
if~$i$ knows that there is an alternative strategy $s_i'$ which, for all joint
strategies of the other players, is strictly preferred.
Furthermore, after iteration $\ell+1$, $i$ knows that $s_i$ is strictly dominated
if~$i$ knows that there is an alternative strategy $s_i'$ such that,
for all joint strategies $s_{-i}$ of the other players,
either $s_i'$ is strictly preferred or some strategy $s_j$ in $s_{-i}$ is already
known by player~$j$ to be strictly dominated after iteration~$\ell$.

Note that for $\ell >1$ each $\dominated^\ell(s_i)$ is a formula of
$\langposk$ that contains occurrences of all $\knows j$ operators.
We restrict our attention to formulas $\dominated^\ell(s_i)$
with $\ell \in \{1, \dots, \hat \ell\}$, where $\hat \ell=\sum_{i\in N}\abs{S_i}$.
By their semantics there is some $\ell$ within this range such that
for all $\ell'\geq \ell$, $\dominated^{\ell'}$ is equivalent to $\dominated^\ell$.
To reflect the fact that this can be seen as the outcome of the iteration,
we denote $\dominated^{\hat \ell}$ by $\dominated^\infty$.



We now proceed to the main result of the paper. We prove that the
non-epistemic formulation of iterated elimination of strictly dominated
strategies, as defined in \cref{sec:iter-strat-elim},
coincides with the epistemic formulation of strict dominance.
As mentioned at the beginning of this section, we can assume that $\G(H,\msgs)$ is
defined using the global version of strict dominance, \sdglobal.

\begin{theorem}
  \label{result:alg-equiv-formula}
  For any  game $\G$, hypergraph $H$, set of messages $\msgs$
  truthful with respect to $\G$, and $i\in N$,
  \[
  \G(H,\msgs)_i=\{s_i\in S_i\mid\state\nvDash\dominated^\infty(s_i)\}\mpunct,
  \]
where $V$ is the valuation induced by~$\G$.
\end{theorem}
Note that, as described in Subsection 1.1, we assume common knowledge of
rationality. Which strategies can be (iteratively) eliminated thus
depends only on the agents' knowledge. The eliminations that can be
performed in the epistemic situation arising from a given state of
communication are captured by the right-hand side of the equation
above.

Recall that the definition of the $\dominated^\infty$ relation involves knowledge
operators $K_i$. However, it does not rely on the common knowledge operators $C_A$.
In the course of the proof of the above theorem (in \cref{result:T-A-iff-CK-A}) we show that common knowledge
is in fact implicitly present here. The technical explanation for this fact is that in  
the definition of $\dominated^\infty$ the operators $K_i$ are nested and that in our context
these operators distribute both over conjunction and disjunction (\cref{result:k-disjunction-distributes}).

\begin{movedProof}  
In order to prove \cref{result:alg-equiv-formula}, we need some preparatory steps.

\begin{lemma}
  \label{result:dom-inner-ck}
  For any $\ell\geq1$, $i\in\agents$, $s_i\in S_i$, and state \state,
  \begin{align*}
    \state&\vDash\dominated^{\ell+1}(s_i)\\
    \iff\state&\vDash\bigvee_{s_i'\in S_i}\bigwedge_{s_{-i}\in S_{-i}}
    [(\knows i\prf is') \vee \bigvee_{\hyarc:i\in \hyarc\in \overline{H}}\bigvee_{j\in \hyarc\setminus\{i\}}\ck \hyarc\dominated^\ell(s_j) ]\mpunct.
  \end{align*}
\end{lemma}
\begin{proof}
  We have
  \begin{align*}
    &\state\vDash\dominated^{\ell+1}(s_i)\\
    \iff[by definition]
    &\state\vDash\knows i\bigvee_{s_i'\in S_i}\bigwedge_{s_{-i}\in S_{-i}}
    [\prf is' \vee \bigvee_{j\in N\setminus\{i\}}\dominated^\ell(s_j) ]\\
    \ifff{by \cref{result:k-disjunction-distributes}}
    &\state\vDash\bigvee_{s_i'\in S_i}\bigwedge_{s_{-i}\in S_{-i}}
    [(\knows i\prf is') \vee \bigvee_{j\in N\setminus\{i\}}\knows i\dominated^\ell(s_j) ]\\
    \iff[]&\state\vDash\bigvee_{s_i'\in S_i}\bigwedge_{s_{-i}\in S_{-i}}
    [(\knows i\prf is') \vee \bigvee_{\hyarc:i\in \hyarc\in \overline{H}}\bigvee_{j\in \hyarc\setminus\{i\}}\ck \hyarc\dominated^\ell(s_j) ]\mpunct.
  \end{align*}
  To see that the downwards implication of the last step holds,
  note that $\dominated^\ell(s_j)=\knows j\varphi$ for appropriate $\varphi$.
  With \cref{result:knowledge-chain-phi-equiv-ck}, $\knows i\knows j\varphi$ implies $\ck{\{i,j\}}\varphi$.
  With an induction starting from \cref{result:knowledge-chain-fact-equiv-msgs-equiv-ck}
  and using \cref{result:ck-disjunction-distributes},
  this implies that there must be messages in~$\msgs$ jointly observed by~$i$ and~$j$
  that entail~$\varphi$.
  Each of these messages must have been sent to some $A\in H$,
  and so all messages have been observed by some $A\in\overline{H}$ with $i,j\in A$.
\end{proof}

\begin{lemma}
  \label{result:T-A-iff-CK-A}
  For any $\ell\geq 1$, $i\in \hyarc\in \overline{H}$, $s_i\in S_i$, and state \state,
  \begin{mainclaim}
    $s_i\not\in (T_{\hyarc,\msgs}^\ell)_i$ iff $\state\vDash\ck \hyarc\dominated^\ell(s_i)$.
  \end{mainclaim}
\end{lemma}
\begin{proof}
  By induction on $\ell$.
  The base case follows straightforwardly from the definitions.
  Now assume the claim holds for $\ell$.
  Then, focusing on the interesting case where $\hyarc\neq\{i\}$,
  we have the following chain of equivalences:
  \begin{align*}
    &s_i\not\in (T_{\hyarc,\msgs}^{\ell+1})_i\\
    \iff[by definition]
    &s_i\not\in (T_{\hyarc,\msgs}^\ell)_i\mvee\neg\sdglobal_{\hyarc,\msgs}(s_i,T_{\hyarc,\msgs}^\ell)\\
    \iff[by contractivity of \sdglobal]
    &\neg\sdglobal_{\hyarc,\msgs}(s_i,T_{\hyarc,\msgs}^\ell)\\
    \iff[by definition]
    &\mexists s'_i \in S_i \: \mforall s_{-i} \in (T_{\hyarc,\msgs}^\ell)_{-i} \:
    \msgs\restr_\hyarc\vDash\prf is'\\ 
    \iff[]
    &\mexists s'_i \in S_i \: \mforall s_{-i} \in S_{-i} \: [
    \begin{aligned}[t]
      &\msgs\restr_\hyarc\vDash\prf is'\mvee\\
      &s_{-i}\not\in (T_{\hyarc,\msgs}^\ell)_{-i} ]
    \end{aligned}\\
    \iff[]
    &\mexists s'_i \in S_i \: \mforall s_{-i} \in S_{-i} \: [
    \begin{aligned}[t]
      &\msgs\restr_\hyarc\vDash\prf is'\mvee\\
      &\mexists j\in \hyarc\setminus\{i\} \: s_j\not\in (T_{\hyarc,\msgs}^\ell)_j ]
    \end{aligned}\\
    \iff[by induction hypothesis]
    &\mexists s'_i \in S_i \: \mforall s_{-i} \in S_{-i} \: [
    \begin{aligned}[t]
      &\msgs\restr_\hyarc\vDash\prf is'\mvee\\
      &\mexists j\in \hyarc\setminus\{i\} \: \state\vDash\ck \hyarc\dominated^\ell(s_j) ]
    \end{aligned}\\
    \ifff{by \cref{result:knowledge-chain-fact-equiv-msgs-equiv-ck}}
    &\mexists s'_i \in S_i \: \mforall s_{-i} \in S_{-i} \: [
    \begin{aligned}[t]
      &\state\vDash\ck \hyarc\prf is'\mvee\\
      &\mexists j\in \hyarc\setminus\{i\} \: \state\vDash\ck \hyarc\dominated^\ell(s_j) ]
    \end{aligned}\\
    \iff[]
    &\state\vDash\bigvee_{s'_i\in S_i}\bigwedge_{s_{-i}\in S_{-1}}
    [ \ck \hyarc\prf is'\vee
    \bigvee_{j\in \hyarc\setminus\{i\}}\ck \hyarc\dominated^\ell(s_j) ]\\
    \ifff{by \cref{result:ck-disjunction-distributes}}
    &\state\vDash\ck \hyarc\bigvee_{s'_i\in S_i}\bigwedge_{s_{-i}\in S_{-1}}
    [ \prf is'\vee
    \bigvee_{j\in \hyarc\setminus\{i\}}\dominated^\ell(s_j) ] \\
     \iff[by definition of $C_A$]
     &\state\vDash\ck \hyarc K_i \bigvee_{s'_i\in S_i}\bigwedge_{s_{-i}\in S_{-1}}
     [ \prf is'\vee
     \bigvee_{j\in \hyarc\setminus\{i\}}\dominated^\ell(s_j) ] \\
    \iff[by definition of $\dominated^{\ell + 1}(\cdot)$]
&\state\vDash\ck \hyarc\dominated^{\ell+1}(s_i)\mpunct.
  \end{align*}
  \qedhack
\end{proof}

We are now ready to prove the main result.

\begin{proof}[\proofof\cref{result:alg-equiv-formula}]
  Let
  \[
  S' := \textstyle\bigcap_{\hyarc:i\in \hyarc\in \overline{H}}T_{\hyarc,\msgs}^\infty.
  \]

  We have:
  \begin{align*}
    &s_i\not\in \G(H,\msgs)_i\\
    \iff[by definition]
    &s_i\not\in (T_{\{i\},\msgs}(S'))_i\\
    \iff[]&\neg\sdglobal_{\{i\},\msgs}(s_i,S')\\
    \iff[]&\mexists s'_i \in S_i \: \mforall s_{-i} \in S'_{-i} \:
    \prf is'\\
    \iff[]&\mexists s'_i \in S_i \: \mforall s_{-i} \in S_{-i} \: (\prf is'\mvee s_{-i}\not\in S'_{-i} )\\
    \iff[]&\mexists s'_i \in S_i \: \mforall s_{-i} \in S_{-i} \: (
    \begin{aligned}[t]
      &\prf is'\mvee\\
      &\mexists \hyarc:i\in \hyarc\in \overline{H} \: s_{-i}\not\in (T_{\hyarc,\msgs}^\infty)_{-i} )
    \end{aligned}\\
    \iff[]&\mexists s'_i \in S_i \: \mforall s_{-i} \in S_{-i} \: (
    \begin{aligned}[t]
      &\prf is'\mvee\\
      &\mexists \hyarc:i\in \hyarc\in \overline{H} \: \mexists j\in \hyarc\setminus\{i\} : s_j\not\in (T_{\hyarc,\msgs}^\infty)_j )
    \end{aligned}\\
    \ifff{by \cref{result:T-A-iff-CK-A}}
    &\mexists s'_i \in S_i \: \mforall s_{-i} \in S_{-i} \: (
    \begin{aligned}[t]
      &\prf is'\mvee\\
      &\state\vDash\bigvee_{\hyarc:i\in \hyarc\in \overline{H}}\bigvee_{j\in \hyarc\setminus\{i\}}\ck \hyarc\dominated^\infty(s_j) )
    \end{aligned}\\
    \iff[since $\prf is'\in\atoms_i$]
    &\mexists s'_i \in S_i \: \mforall s_{-i} \in S_{-i} \: (
    \begin{aligned}[t]
      &\state\vDash\knows i\prf is'\mvee\\
      &\state\vDash\bigvee_{\hyarc:i\in \hyarc\in \overline{H}}\bigvee_{j\in \hyarc\setminus\{i\}}\ck \hyarc\dominated^\infty(s_j) )
    \end{aligned}\\
    \iff[]&\state\vDash\bigvee_{s'_i \in S_i}\bigwedge_{s_{-i} \in S_{-i}}
    [(\knows i\prf is')\vee
    \bigvee_{\hyarc:i\in \hyarc\in \overline{H}}\bigvee_{j\in \hyarc\setminus\{i\}}\ck \hyarc\dominated^\infty(s_j)]\\
    \ifff{by \cref{result:dom-inner-ck}}
    &\state\vDash\dominated^\infty(s_i)\mpunct.\qedhere
  \end{align*}
  \qedhack
\end{proof}
\end{movedProof}

\section{Concluding remarks}
\label{sec:conclusions-stratelim}

We studied normal-form games in the presence of interaction
structures.  We assumed that initially the players know only their own
preferences, and that they can truthfully communicate information
about their own preferences within their parts of the interaction
structure.  This allowed us to analyze the consequences of locality,
formalized by means of an interaction structure, on the outcome of the
iterated elimination of strictly dominated strategies.
To this end we appropriately adapted the framework introduced in \cite{apt_knowledge_2009}
and showed that in any given state of communication
this outcome can be described by means of epistemic analysis.

\subsection{Possible extensions}
\label{sec:possible-extensions-stratelim}

It would be interesting to extend the analysis here presented in a number of ways, by:
\begin{itemize}
\item allowing players to send information about the preferences of other players
  that they learned through interaction.

\item allowing other forms of messages,
  for example, messages containing information that a strategy has been eliminated,
  or containing epistemic statements, such as knowing that some strategy of
\emph{another player} has been eliminated;

\item considering formation or evolution of interaction structures,
  given strategic advantages of certain interaction structures over others,

\item (suggested by one of the referees) considering a set up in which
each admitted group of agents uses a correlation device to synchronize
their actions,

\item considering strategic aspects of communication,
  even if truthfulness is required (should one send some piece of information or not?).

\end{itemize}
The last point is discussed further in the subsection below.

We are currently working on a distributed implementation
of the aspects presented in this paper,
which we plan to extend and evaluate in a separate work.
A preliminary description is contained in~\cite{witzel_knowledge_2009}.
The objective is to present it in a form of an agent-oriented program
as suggested by \citet{shoham_agent-oriented_1993},
where the agents' motivational attitudes are based on their payoffs.

Finally, let us mention that in \cite{apt_knowledge_2009} we studied
an abstract (so non-game theoretic) setting in which players send
messages that inform a group about some atomic fact that a player
knows or has learned. We clarified there, among other things, under
what conditions common knowledge of the underlying hypergraph matters.
The framework there considered could be generalized by allowing
players to jointly arrive at some conclusions using their background
theories, by interaction through messages sent to groups. From this
perspective the form of IESDS studied here could be seen as an
instance of such a conclusion. Because of the form of
allowed messages and background knowledge, this study would differ
from the line of research pursued by \citet{fagin_reasoning_1995},
where the effects of communication are considered in the framework of
distributed systems.

\subsection{Strategic communication}
\label{sec:stra}

Among the topics for future research especially incorporation of
strategic communication into our framework is an interesting
challenge.  Note that we do \emph{not} examine here strategic or
normative aspects of the \emph{communication}.  In fact, we do not
allow players to lie and do not even examine \emph{why} they
communicate or \emph{what} they should truthfully communicate to
maximize their utilities.  Rather, we examine what happens \emph{if}
they do communicate, assuming that they are truthful, rational and
have reasoning powers.

To justify this focus, it is helpful to realize that in some settings
strategic aspects of communication are not relevant.  One possibility
is when communication is not a deliberate act, but rather occurs
through observation of somebody's behavior.
Such communication is certainly more difficult
to manipulate and more laborious to fake than mere words.  In a sense
it is inherently credible, and research in social learning argues
along similar lines~\cite[Ch.~3]{chamley_rational_2004}.


In the setting of artificial agents communicating by 
messages, to view communication as something non-deliberate is more
problematic.  Here, ignoring strategic aspects of communication can be
interpreted as bounds on the players' rationality or reasoning
capabilities---they simply lack the capabilities to deal with all the
consequences of such an inherently rich phenomenon as communication.

In general, strategic communication is a research topic on its own,
with controversial discussions (see, e.g., \cite{sally_can_2005}) and
many questions open.  \citet{crawford_strategic_1982} have
considered the topic in a probabilistic setting, and
\citet{farrell_cheap_1996} have looked at related issues under the
notion of \dfnless{cheap talk}.  Within epistemic logic,
formalizations of the information content of strategic communication
have been suggested, e.g., by \citet{gerbrandy_communication_2007}.

\subsection{Dynamics of group communication}

Furthermore, the subject of group communication has other
connections to epistemic logic:
the changes in knowledge that are brought about by various kinds of actions,
including some forms of communication, have been extensively examined
in so-called
\emph{dynamic} epistemic logic.
\citet{BMS} is an important early paper in this field,
introducing an operation on epistemic models
that allows for the rigorous analysis of a very broad class of epistemic events.
That class includes the announcements to subgroups that we in effect
consider here.
\citet{BMS} study logics in which there are, in addition to modal
operators for knowledge and common knowledge, operators $[a]\varphi$,
meaning that after action $a$, $\varphi$ holds.
Our interest here is not directly related to these concerns.  We do
not have (or need) a logic for describing states, and we consider a
specific communication protocol that is formalized by our notion of
\emph{interaction structure}.  We use this to induce a solution
concept, whose epistemic foundation we then study. 
On the other hand a dynamic epistemic logic might be helpful to
describe precisely
the dynamics of the strategy elimination process, a topic that we do
not consider here.



\section*{Acknowledgements}
\label{sec:acknowledgements}

We thank Rohit Parikh and Willemien Kets
for discussion and helpful suggestions.
We also thank three anonymous referees for their very constructive comments.

\bibliographystyle{abbrvnat}
\bibliography{mye,misc,all}

\ifx\Newassociation\undefined\else\Closesolutionfile{movedProofs}\fi

\ifx\fullversion\undefined
\else
\appendix
\section{Omitted Proofs}
\label{sec:omitted-proofs}
\ifx\Newassociation\undefined\begin{MovedProof}{[subsection][1][3]3.1}
\begin{proof}[\proofof\cref{thm:hyper}]
We first prove the corresponding claim for the global version of strict dominance, \sdglobal,
introduced before \cref{lem:l-equals-g}.

Consider $\group \subseteq \group'\subseteq N$.
By definition, this implies that
for all restrictions $\G'$ we have $T_{\group'}(\G') \subseteq T_\group(\G')$.
Since $\sdglobal$ is monotonic, so is the operator $T_C$ for all $C \subseteq N$.
Hence by a straightforward induction $T_N^{\infty} \subseteq T_\group^{\infty}$ for all $\group \subseteq N$,
and consequently, for all players $i$,
\begin{equation}
\label{eqn:comminclu}
T_N^{\infty} \subseteq \textstyle\bigcap_{\hyarc: i \in \hyarc \in H} T^{\infty}_\hyarc\mpunct.
\end{equation}
Hence, for all $i \in N$,
\[
T_N^{\infty} = T_{\{i\}}(T_N^{\infty}) \subseteq T_{\{i\}}(\textstyle\bigcap_{\hyarc: i \in \hyarc \in H} T^{\infty}_\hyarc),
\]
where the inclusion holds by the monotonicity of $T_{\{i\}}$.
Consequently $T^{\infty}_N \subseteq \G(H)$.

We now prove the original claim for the local version of strict dominance.
We need to distinguish the $T_C$ operator for $\sdlocal$ and $\sdglobal$. In the former
case we write $T_{C, l}$ and in the latter case $T_{C, g}$. The reason that we use the latter
operators is that they are monotonic and closely related to the former operators.
As a consequence of \cref{lem:l-equals-g}, $T^{\infty}_{C, l} = T^{\infty}_{C, g}$.
Now fix an arbitrary $i \in N$, then
\[
\textstyle\bigcap_{\hyarc: i \in \hyarc \in H} T^{\infty}_{\hyarc,g} = \textstyle\bigcap_{\hyarc: i \in \hyarc \in H} T^{\infty}_{\hyarc,l}\mpunct,
\]
and by \eqref{eqn:comminclu} for $\sdglobal$, $T^{\infty}_{N, g} \subseteq \bigcap_{\hyarc: i \in \hyarc \in H} T^{\infty}_{\hyarc,g}$, so
\begin{equation}
  \label{equ:lg}
T^{\infty}_{N, l} = T^{\infty}_{N, g} \subseteq \textstyle\bigcap_{\hyarc: i \in \hyarc \in H} T^{\infty}_{\hyarc,l}\mpunct.
\end{equation}
Further, we have
$T^{\infty}_{N, l} = T^{\infty}_{N, g}$ and
$T_{N, g}^{\infty} = T_{\{i\},g}(T_{N, g}^{\infty})$, so
$T_{N, l}^{\infty} = T_{\{i\},g}(T_{N, l}^{\infty})$.
Hence, by \eqref{equ:lg} and monotonicity of $T_{\{i\},g}$,
\[
T^{\infty}_{N, l} = T_{\{i\},g}(T_{N, l}^{\infty}) \subseteq T_{\{i\},g}(\textstyle\bigcap_{\hyarc: i \in \hyarc \in H} T^{\infty}_{\hyarc,l})\mpunct.
\]
Also, for all $i \in N$ and all restrictions $\G'$ we have, by definition,
\[T_{\{i\}, g}(\G') \subseteq T_{\{i\}, l}(\G')\mpunct,\]
so by the last inclusion
\[
T^{\infty}_{N, l} \subseteq T_{\{i\},l}(\textstyle\bigcap_{\hyarc: i \in \hyarc \in H} T^{\infty}_{\hyarc,l})\mpunct.
\]
Consequently, $T^{\infty}_{N,l} \subseteq \G(H)$, as desired.

\end{proof}
\end{MovedProof}
\begin{MovedProof}{[subsection][2][4]4.2}
\begin{proof}[\proofof\cref{lem:l-equals-g}]
  We show that for all $k\geq0$ each globally (strictly) dominated
  strategy is also locally dominated in $\G^k$.  Together with the
  straightforward fact that local dominance implies global dominance,
  this proves the desired equivalence.

  Formally, the claim is thus
  that for all $k\geq0$, $s_i\in S_i$ and $s_i'\in \G^k_i$ such that
  $\pref{i}{\G^k}{s}{s'}$, there is $s_i''\in \G^k_i$ with
  $\pref{i}{\G^k}{s''}{s'}$.
  To show this we prove by induction that for all $k\geq0$ and $s_i\in
  S_i$, there is $s_i''\in \G^k_i$ such that
  $\prefeq{i}{\G^k}{s''}{s}$, from which the claim follows since
$\prefeq{i}{\G^k}{s''}{s}$ and $\pref{i}{\G^k}{s}{s'}$ imply $\pref{i}{\G^k}{s''}{s'}$.

This claim clearly holds for $k = 0$.  Now assume the statement holds
for some $k$ and fix $s_i\in S_i$.  Choose some $s_i'\in S_i$ that
is $\preferredeq{\G^k_{-i}}{}{}$-maximal among the elements of $S_i$.
By the induction hypothesis there is $s_i''\in \G^k_i$ such that
$\prefeq{i}{\G^k}{s''}{s'}$.
So also $s_i''$ is $\preferredeq{\G^k_{-i}}{}{}$-maximal among the elements of $S_i$.
Hence $s_i'' \in (\sdglobal(\G^k))_i$.

  From $\sdglobal(\G^k)\subseteq \G^{k+1}$ we now obtain $s_i''\in \G^{k+1}_i$.
  From $\G^{k+1}\subseteq \G^k$ and $\prefeq{i}{\G^k}{s''}{s'}$, we obtain $\prefeq{i}{\G^{k+1}}{s''}{s'}$,
  and by the maximality of $s_i'$, we also have $\prefeq{i}{\G^{k+1}}{s'}{s}$.
  Thus, $s_i''\in \G^{k+1}_i$ and $\prefeq{i}{\G^{k+1}}{s''}{s}$, which concludes the proof of the induction step.
\end{proof}
\end{MovedProof}
\begin{MovedProof}{[subsection][2][4]4.2}
In order to prove \cref{thm:lg}, we need an auxiliary lemma dealing with operators in a general setting.
Given $Y \in D$ and an operator $T$  on a finite lattice $(D, \subseteq)$,
we denote by $T_Y$ the following operator:
\[
T_Y(X) := T(X) \cup (X \cap Y).
\]

\begin{numberednote}\label{not:T}
  If the $T$ operator is contracting, then so is $T_Y$.
\end{numberednote}

\begin{lemma}\label{lem:TU}
  Suppose that $T$ and $U$ are operators on a finite lattice $(D,
  \subseteq)$ such that $T$ is monotonic and contracting.  Then $
  T_{T^{\infty} \cap U^{\infty}}^{\infty} (U^{\infty}) = T^{\infty}
  \cap U^{\infty} $.
\end{lemma}

Informally, this claim states that the combined effect of independent limit
iterations of $T$ and $U$ can be modelled by `serial' limit iterations of
$T$ and $U$, provided the operator $T$ is modified to an appropriate
$T_Y$ form.
\medskip

\begin{proof}

Denote for brevity $T^{\infty} \cap U^{\infty}$ by $Y$.
First we prove by induction that for all $k \geq 0$
\[
Y \subseteq T_{Y}^{k} (U^{\infty}).
\]
The claim clearly holds for $k = 0$.
Suppose it holds for some $k \geq 0$. Then by the induction hypothesis
\[
T_{Y}^{k+1}(U^{\infty}) = T(T_{Y}^{k}(U^{\infty})) \cup (T_{Y}^{k}(U^{\infty}) \cap Y) \supseteq Y.
\]
Hence
\begin{equation}
  \label{equ:TU}
Y \subseteq T_{Y}^{\infty} (U^{\infty}).
\end{equation}

To prove the converse implication we show by induction that for all $k \geq 0$
\[
T_{Y}^{k}(U^{\infty}) \subseteq T^{k}.
\]
The claim clearly holds for $k = 0$.
Suppose it holds for some $k \geq 0$. Then by the induction hypothesis
and the monotonicity of $T$
\[
T_{Y}^{k+1}(U^{\infty}) = T_Y(T_{Y}^{k}(U^{\infty})) \subseteq T_Y(T^{k}) =
T^{k+1} \cup (T^{k} \cap Y) \subseteq T^{k+1} \cup T^{\infty}  \subseteq T^{k+1}.
\]
Hence
\begin{equation}
  \label{equ:T}
T_{Y}^{\infty} (U^{\infty}) \subseteq T^{\infty}.
\end{equation}

Next, by \cref{not:T} the operator $T_{Y}$ is contracting, so
\begin{equation}
  \label{equ:U}
T_{Y}^{\infty} (U^{\infty}) \subseteq U^{\infty}.
\end{equation}

Now the claim follows by \eqref{equ:TU}, \eqref{equ:T} and \eqref{equ:U}.
\end{proof}

\begin{proof}[\proofof\cref{thm:lg}]

Recall that $\overline{H}$ denotes the closure of~$H$ under non-empty intersection.
Fix an interaction structure $H$, a set of messages $\msgs$ and $i \in
N$.  Assume for simplicity that the set $\{A \mid i
\in A \in \overline{H}\}$ has exactly two elements, say, $B_i$ and
$C_i$.  To deal with the arbitrary situation \cref{lem:TU} needs
to be generalized to an arbitrary number of operators. Such a
generalization is straightforward and omitted.

Let now
\[
\begin{array}{l}
\G_1 := (S_1, \ldots, S_n),  \\
\G_2 := T^{\infty}_{B_i, \msgs}(\G_1), \\
\G_3 := \hat{T}^{\infty}_{C_i, \msgs}(\G_2), \\
\G_4 := T_{\{i\}}(\G_3),
\end{array}
\]
where
\[
\hat{T}_{C_i, \msgs}(\G) := T_{C_i, \msgs}(\G) \cup (\G \cap T^{\infty}_{B_i, \msgs} \cap T^{\infty}_{C_i, \msgs}).
\]

Now, recall that
 \[
 \G(H,\msgs)_i := \left(T_{\{i\},\msgs}\left(\textstyle\bigcap_{\hyarc: i \in \hyarc \in\overline{H}} T^{\infty}_{\hyarc,\msgs}\right)\right)_i\mpunct.
 \]
By \cref{lem:TU}, $\G_3 = T^{\infty}_{B_i, \msgs} \cap
T^{\infty}_{C_i, \msgs}$, so $(\G_4)_i$ is $\G(H,\msgs)_i$, the $i$th component of $\G(H,\msgs)$.
\Cref{not:T} ensures that each of the operators
$T_{C_i, \msgs}, T_{B_i, \msgs}$ and $\hat{T}_{C_i, \msgs}$
is contracting. Moreover, $\sdglobal$ removes (weakly) more strategies than each of them, so
the sequence of restrictions
\[
\G_1, \ T_{B_i, \msgs}(\G_1), \ T^{2}_{B_i, \msgs}(\G_1), \ldots, \G_2, \
\hat{T}_{C_i, \msgs}(\G_2), \ \hat{T}^{2}_{C_i, \msgs}(\G_2), \dots, \G_3, \
\G_4
\]
satisfies the conditions of \cref{lem:l-equals-g}.
By \cref{lem:l-equals-g} we also obtain the same restriction $\G_3$ when in the
definition of the $T_{B_i, \msgs}$ and $T_{C_i, \msgs}$ operators we use
$\sdlocal$ instead of $\sdglobal$.  So the $i$th component
$\G(H,\msgs)_i$
of
$\G(H,\msgs)$
is the
same when in the definitions of the $T_{B_i, \msgs}$, $T_{C_i, \msgs}$ and
$T_{\{i\}}$ operators we use $sd^{\ell}$ instead of $sd^g$.

This concludes the proof.
\end{proof}
\end{MovedProof}
\begin{MovedProof}{[subsection][2][4]4.2}
In order to prove \cref{result:alg-equiv-formula}, we need some preparatory steps.

\begin{lemma}
  \label{result:dom-inner-ck}
  For any $\ell\geq1$, $i\in\agents$, $s_i\in S_i$, and state \state,
  \begin{align*}
    \state&\vDash\dominated^{\ell+1}(s_i)\\
    \iff\state&\vDash\bigvee_{s_i'\in S_i}\bigwedge_{s_{-i}\in S_{-i}}
    [(\knows i\prf is') \vee \bigvee_{\hyarc:i\in \hyarc\in \overline{H}}\bigvee_{j\in \hyarc\setminus\{i\}}\ck \hyarc\dominated^\ell(s_j) ]\mpunct.
  \end{align*}
\end{lemma}
\begin{proof}
  We have
  \begin{align*}
    &\state\vDash\dominated^{\ell+1}(s_i)\\
    \iff[by definition]
    &\state\vDash\knows i\bigvee_{s_i'\in S_i}\bigwedge_{s_{-i}\in S_{-i}}
    [\prf is' \vee \bigvee_{j\in N\setminus\{i\}}\dominated^\ell(s_j) ]\\
    \ifff{by \cref{result:k-disjunction-distributes}}
    &\state\vDash\bigvee_{s_i'\in S_i}\bigwedge_{s_{-i}\in S_{-i}}
    [(\knows i\prf is') \vee \bigvee_{j\in N\setminus\{i\}}\knows i\dominated^\ell(s_j) ]\\
    \iff[]&\state\vDash\bigvee_{s_i'\in S_i}\bigwedge_{s_{-i}\in S_{-i}}
    [(\knows i\prf is') \vee \bigvee_{\hyarc:i\in \hyarc\in \overline{H}}\bigvee_{j\in \hyarc\setminus\{i\}}\ck \hyarc\dominated^\ell(s_j) ]\mpunct.
  \end{align*}
  To see that the downwards implication of the last step holds,
  note that $\dominated^\ell(s_j)=\knows j\varphi$ for appropriate $\varphi$.
  With \cref{result:knowledge-chain-phi-equiv-ck}, $\knows i\knows j\varphi$ implies $\ck{\{i,j\}}\varphi$.
  With an induction starting from \cref{result:knowledge-chain-fact-equiv-msgs-equiv-ck}
  and using \cref{result:ck-disjunction-distributes},
  this implies that there must be messages in~$\msgs$ jointly observed by~$i$ and~$j$
  that entail~$\varphi$.
  Each of these messages must have been sent to some $A\in H$,
  and so all messages have been observed by some $A\in\overline{H}$ with $i,j\in A$.
\end{proof}

\begin{lemma}
  \label{result:T-A-iff-CK-A}
  For any $\ell\geq 1$, $i\in \hyarc\in \overline{H}$, $s_i\in S_i$, and state \state,
  \begin{mainclaim}
    $s_i\not\in (T_{\hyarc,\msgs}^\ell)_i$ iff $\state\vDash\ck \hyarc\dominated^\ell(s_i)$.
  \end{mainclaim}
\end{lemma}
\begin{proof}
  By induction on $\ell$.
  The base case follows straightforwardly from the definitions.
  Now assume the claim holds for $\ell$.
  Then, focusing on the interesting case where $\hyarc\neq\{i\}$,
  we have the following chain of equivalences:
  \begin{align*}
    &s_i\not\in (T_{\hyarc,\msgs}^{\ell+1})_i\\
    \iff[by definition]
    &s_i\not\in (T_{\hyarc,\msgs}^\ell)_i\mvee\neg\sdglobal_{\hyarc,\msgs}(s_i,T_{\hyarc,\msgs}^\ell)\\
    \iff[by contractivity of \sdglobal]
    &\neg\sdglobal_{\hyarc,\msgs}(s_i,T_{\hyarc,\msgs}^\ell)\\
    \iff[by definition]
    &\mexists s'_i \in S_i \: \mforall s_{-i} \in (T_{\hyarc,\msgs}^\ell)_{-i} \:
    \msgs\restr_\hyarc\vDash\prf is'\\ 
    \iff[]
    &\mexists s'_i \in S_i \: \mforall s_{-i} \in S_{-i} \: [
    \begin{aligned}[t]
      &\msgs\restr_\hyarc\vDash\prf is'\mvee\\
      &s_{-i}\not\in (T_{\hyarc,\msgs}^\ell)_{-i} ]
    \end{aligned}\\
    \iff[]
    &\mexists s'_i \in S_i \: \mforall s_{-i} \in S_{-i} \: [
    \begin{aligned}[t]
      &\msgs\restr_\hyarc\vDash\prf is'\mvee\\
      &\mexists j\in \hyarc\setminus\{i\} \: s_j\not\in (T_{\hyarc,\msgs}^\ell)_j ]
    \end{aligned}\\
    \iff[by induction hypothesis]
    &\mexists s'_i \in S_i \: \mforall s_{-i} \in S_{-i} \: [
    \begin{aligned}[t]
      &\msgs\restr_\hyarc\vDash\prf is'\mvee\\
      &\mexists j\in \hyarc\setminus\{i\} \: \state\vDash\ck \hyarc\dominated^\ell(s_j) ]
    \end{aligned}\\
    \ifff{by \cref{result:knowledge-chain-fact-equiv-msgs-equiv-ck}}
    &\mexists s'_i \in S_i \: \mforall s_{-i} \in S_{-i} \: [
    \begin{aligned}[t]
      &\state\vDash\ck \hyarc\prf is'\mvee\\
      &\mexists j\in \hyarc\setminus\{i\} \: \state\vDash\ck \hyarc\dominated^\ell(s_j) ]
    \end{aligned}\\
    \iff[]
    &\state\vDash\bigvee_{s'_i\in S_i}\bigwedge_{s_{-i}\in S_{-1}}
    [ \ck \hyarc\prf is'\vee
    \bigvee_{j\in \hyarc\setminus\{i\}}\ck \hyarc\dominated^\ell(s_j) ]\\
    \ifff{by \cref{result:ck-disjunction-distributes}}
    &\state\vDash\ck \hyarc\bigvee_{s'_i\in S_i}\bigwedge_{s_{-i}\in S_{-1}}
    [ \prf is'\vee
    \bigvee_{j\in \hyarc\setminus\{i\}}\dominated^\ell(s_j) ] \\
     \iff[by definition of $C_A$]
     &\state\vDash\ck \hyarc K_i \bigvee_{s'_i\in S_i}\bigwedge_{s_{-i}\in S_{-1}}
     [ \prf is'\vee
     \bigvee_{j\in \hyarc\setminus\{i\}}\dominated^\ell(s_j) ] \\
    \iff[by definition of $\dominated^{\ell + 1}(\cdot)$]
&\state\vDash\ck \hyarc\dominated^{\ell+1}(s_i)\mpunct.
  \end{align*}
  \qedhack
\end{proof}

We are now ready to prove the main result.

\begin{proof}[\proofof\cref{result:alg-equiv-formula}]
  Let
  \[
  S' := \textstyle\bigcap_{\hyarc:i\in \hyarc\in \overline{H}}T_{\hyarc,\msgs}^\infty.
  \]

  We have:
  \begin{align*}
    &s_i\not\in \G(H,\msgs)_i\\
    \iff[by definition]
    &s_i\not\in (T_{\{i\},\msgs}(S'))_i\\
    \iff[]&\neg\sdglobal_{\{i\},\msgs}(s_i,S')\\
    \iff[]&\mexists s'_i \in S_i \: \mforall s_{-i} \in S'_{-i} \:
    \prf is'\\
    \iff[]&\mexists s'_i \in S_i \: \mforall s_{-i} \in S_{-i} \: (\prf is'\mvee s_{-i}\not\in S'_{-i} )\\
    \iff[]&\mexists s'_i \in S_i \: \mforall s_{-i} \in S_{-i} \: (
    \begin{aligned}[t]
      &\prf is'\mvee\\
      &\mexists \hyarc:i\in \hyarc\in \overline{H} \: s_{-i}\not\in (T_{\hyarc,\msgs}^\infty)_{-i} )
    \end{aligned}\\
    \iff[]&\mexists s'_i \in S_i \: \mforall s_{-i} \in S_{-i} \: (
    \begin{aligned}[t]
      &\prf is'\mvee\\
      &\mexists \hyarc:i\in \hyarc\in \overline{H} \: \mexists j\in \hyarc\setminus\{i\} : s_j\not\in (T_{\hyarc,\msgs}^\infty)_j )
    \end{aligned}\\
    \ifff{by \cref{result:T-A-iff-CK-A}}
    &\mexists s'_i \in S_i \: \mforall s_{-i} \in S_{-i} \: (
    \begin{aligned}[t]
      &\prf is'\mvee\\
      &\state\vDash\bigvee_{\hyarc:i\in \hyarc\in \overline{H}}\bigvee_{j\in \hyarc\setminus\{i\}}\ck \hyarc\dominated^\infty(s_j) )
    \end{aligned}\\
    \iff[since $\prf is'\in\atoms_i$]
    &\mexists s'_i \in S_i \: \mforall s_{-i} \in S_{-i} \: (
    \begin{aligned}[t]
      &\state\vDash\knows i\prf is'\mvee\\
      &\state\vDash\bigvee_{\hyarc:i\in \hyarc\in \overline{H}}\bigvee_{j\in \hyarc\setminus\{i\}}\ck \hyarc\dominated^\infty(s_j) )
    \end{aligned}\\
    \iff[]&\state\vDash\bigvee_{s'_i \in S_i}\bigwedge_{s_{-i} \in S_{-i}}
    [(\knows i\prf is')\vee
    \bigvee_{\hyarc:i\in \hyarc\in \overline{H}}\bigvee_{j\in \hyarc\setminus\{i\}}\ck \hyarc\dominated^\infty(s_j)]\\
    \ifff{by \cref{result:dom-inner-ck}}
    &\state\vDash\dominated^\infty(s_i)\mpunct.\qedhere
  \end{align*}
  \qedhack
\end{proof}
\end{MovedProof}
\else\Readsolutionfile{movedProofs}\fi
\fi

\end{document}